\documentclass[prx,reprint,floatfix,letterpaper,twocolumn,nofootinbib,showpacs,longbibliography]{revtex4-1}
\usepackage[caption=false]{subfig}
\usepackage{graphicx} 
\usepackage{epstopdf}

\usepackage{array} 
\usepackage{verbatim} 
\usepackage{amsmath,amsfonts,amssymb,amscd}
\usepackage{amsthm}
\usepackage{tabularx}
\usepackage{stmaryrd}
\usepackage{enumerate}
\usepackage[ruled]{algorithm2e}

\usepackage{hyperref}
\hypersetup{
    bookmarksnumbered=true, 
    unicode=false, 
    pdfstartview={FitH}, 
    pdftitle={}, 
    pdfauthor={}, 
    pdfsubject={}, 
    pdfcreator={}, 
    pdfproducer={}, 
    pdfkeywords={}, 
    pdfnewwindow=true, 
    colorlinks=true, 
    linkcolor=blue, 
    citecolor=blue, 
    filecolor=blue, 
    urlcolor=blue 
}

\theoremstyle{plain}
\newtheorem{thm}{Theorem}[section]
\newtheorem{cor}[thm]{Corollary}
\newtheorem{lem}[thm]{Lemma}

\theoremstyle{definition}

\newcommand{\eq}[1]{(\hyperref[eq:#1]{\ref*{eq:#1}})}

\renewcommand{\sec}[1]{\hyperref[sec:#1]{Section~\ref*{sec:#1}}}
\newcommand{\thrm}[1]{\hyperref[thm:#1]{Theorem~\ref*{thm:#1}}}
\newcommand{\lemm}[1]{\hyperref[lemm:#1]{Lemma~\ref*{lemm:#1}}}
\newcommand{\prop}[1]{\hyperref[prop:#1]{Proposition~\ref*{prop:#1}}}
\newcommand{\corr}[1]{\hyperref[corr:#1]{Corollary~\ref*{corr:#1}}}
\newcommand{\fig}[1]{\hyperref[fig:#1]{Figure~\ref*{fig:#1}}}

\newcommand{\ket}[1]{|#1\rangle}

\usepackage[usenames,dvipsnames]{color}
\usepackage{xcolor}

\DeclareMathAlphabet{\matheu}{U}{eus}{m}{n}

\newcommand{\X}{{\mathbb P}_X}

\newcolumntype{L}[1]{>{\raggedright}p{#1}}
\newcolumntype{C}[1]{>{\centering}p{#1}}
\newcolumntype{R}[1]{>{\raggedleft}p{#1}}
\newcolumntype{D}{>{\centering\arraybackslash}X}

\begin{document}
\title{Universal fault-tolerant gates on concatenated stabilizer codes}
\author{Theodore J. Yoder}
\author{Ryuji Takagi}
\author{Isaac L. Chuang}
\affiliation{Massachusetts Institute of Technology, Cambridge, MA 02139}

\begin{abstract}
It is an oft cited fact that no quantum code can support a set of fault-tolerant logical gates that is both universal and transversal. This no-go theorem is generally responsible for the interest in alternative universality constructions including magic state distillation. Widely overlooked, however, is the possibility of non-transversal, yet still fault-tolerant, gates that work directly on small quantum codes. Here we demonstrate precisely the existence of such gates. In particular, we show how the limits of non-transversality can be overcome by performing rounds of intermediate error-correction to create logical gates on stabilizer codes that use no ancillas other than those required for syndrome measurement. Moreover, the logical gates we construct, the most prominent examples being Toffoli and controlled-controlled-$Z$, often complete universal gate sets on their codes. We detail such universal constructions for the smallest quantum codes, the 5-qubit and 7-qubit codes, and then proceed to generalize the approach. One remarkable result of this generalization is that any nondegenerate stabilizer code with a complete set of fault-tolerant single-qubit Clifford gates has a universal set of fault-tolerant gates. Another is the interaction of logical qubits across \emph{different} stabilizer codes, which, for instance, implies a broadly applicable method of code switching.
\end{abstract}


\maketitle

\section{Introduction}
One of the crucial concepts in error-correcting codes is that of logical circuits -- a set of circuits $\{C_i\}$ that give the ability to carry out a set of operations $\{U_i\}$ directly on encoded data, rather than performing the risky procedure of decoding, applying $U_i$, and re-encoding. In fact, the latter procedure is forbidden if we insist on each $C_i$ being a fault-tolerant logical circuit, for which the failure of any one component in $C_i$ never leads to an uncorrectable error on the encoded data. If, in addition, the set $\{U_i\}$ is universal for the computational model in question (e.g. classical or quantum computation), then the set of logical circuits $\{C_i\}$ is said to be universal, and the error-correcting code in question could in principle be used for all computational purposes without ever needing to decode, an essential ability for quantum computing especially, where decoherence remains the bane of all practical implementations of quantum algorithms.

In the quantum computational model, the one paradigm for designing fault-tolerant logical gates that is most preferred is transversality. A logical circuit is transversal if all physical qubits have interacted with at most one physical qubit from each code block, and if such a design preserves the code space then it is automatically fault-tolerant. Indeed, many quantum codes, such as Steane's 7-qubit code \cite{Steane1996}, are highly regarded exactly because they have important, and perhaps several important, transversal gates.

Unfortunately, it is a well-known theorem \cite{Chen2008,Eastin2009,Zeng2011} that there is no quantum code with a universal set of transversal logical circuits. This fact means other methods must be used to perform universal, fault-tolerant quantum computation. The most common approach is that of magic states \cite{Bravyi2005,Bravyi2012}, encoded ancilla qubits that, combined with available transversal circuits (usually implementing Clifford operations), serve to complete a universal set of logical circuits (usually by implementing a $T$ or Toffoli gate). Ideally, these magic states would be efficiently constructible themselves, but current so-called distillation procedures actually incur large overheads in terms of time and qubits \cite{Fowler2012,Fowler2013}.

It is therefore fortunate that other approaches to bypass the universal-transversal no-go theorem exist. Some involve code switching --- the transversal circuits on two codes together might complete a universal set, encouraging development of a method to exchange data between the two code spaces. Simple procedures have been devised for conversions between specific codes, such as between the 5-qubit and 7-qubit \cite{Hill2013} and between quantum Reed-Muller codes \cite{Anderson2014}. Another workaround for the no-go theorem uses tri-orthogonal subsystem codes \cite{Paetznick2013}, the smallest of which is 15-qubits, to implement a universal, transversal set of gates without code switching, but with additional error-correction on the gauge qubits. Related gauge-fixing techniques are employed in topological color codes \cite{Bombin2015}, where getting locally implementable non-Clifford gates requires jumping up a dimension to 3D \cite{Bombin2016}. Lastly, concatenated coding combines two codes with complementary transversal gate sets into a single large code with universal, yet not transversal, gates \cite{Jochym2014}.

Here we complement the myriad previous approaches to universal fault-tolerance by developing non-transversal, yet still fault-tolerant, circuits implementing logical gates on stabilizer codes. Our approach is based on an under-appreciated trick first used by Knill, Laflamme, and Zurek \cite{Knill1996} to implement a fault-tolerant controlled-$S$ gate on the 7-qubit code. The trick revolves around breaking a non-transversal circuit into fault-tolerant pieces, with error-correction performed in-between to correct errors before they propagate too badly throughout the non-transversal circuit. We show that this trick for creating fault-tolerant logical gates can be significantly generalized to perform other logical gates on other codes and ultimately developed into a procedure we call pieceable fault-tolerance.

As examples of pieceable fault-tolerance, we create a logical CZ gate on the 5-qubit code (Section~\ref{5qubit}) and logical CCZ gates on the 5-qubit (\ref{5qubit}) and 7-qubit (\ref{7qubit}) codes, completing universal sets of gates on those codes. Also notable is that our circuits use no ancillas other than those required for the multiple rounds of error-correction --- in particular, we use no magic-states. Indeed, in Section~\ref{7qubit} we find our construction for the 7-qubit code compares favorably against magic state injection with regards to resources required. All our pieceable circuits are 1-fault-tolerant. That is, any \emph{one} fault does not cause a logical error, as is consistent with both of these small codes being distance three. Through concatenation, these pieceable circuits also possess a fault-tolerance threshold in the usual sense \cite{Knill1996a,Aliferis2005}.

In addition, we provide sufficient conditions for similar pieceable circuits to work on larger codes more generally (Section~\ref{conditions_for_pieceably}). We find that nondegeneracy or, more specifically, not having weight two stabilizers, is sufficient (albeit not necessary) for a code to have a pieceably 1-fault-tolerant gate. Therefore, any nondegenerate stabilizer code with a universal set of fault-tolerant \emph{local} (i.e. single-qubit) Clifford logical gates can be promoted to fault-tolerant universality using our pieceable methods. Compare to magic states, where to achieve the same universality for distance three codes, even with much more overhead, the entire set of logical Cliffords is required \cite{Bravyi2005}. Moreover, we show nondegenerate CSS codes just need \emph{any} fault-tolerant local Clifford to achieve the same universality. Finally, we find that pieceable fault-tolerance can also perform gates between different codes, and thus act as a quite general method of code switching.

Noteworthy ideas in the literature closely related to pieceable fault-tolerance are the code switching example of Hill et.~al.~\cite{Hill2013} and code deformations of Bombin and Martin-Delgado \cite{Bombin2009,Bombin2011}. In the former, a circuit of Clifford gates with error-correction performed after each gate is used to switch between the 5-qubit and 7-qubit codes. In the latter, logical initialization, measurement, and Clifford gates are performed on surface codes by manipulating the geometry of the surface while correcting errors as they arise. Both techniques are similar to pieceable fault-tolerance as the code undergoes transformations to several intermediate codes, each with distance large enough to correct any errors that may have arisen. However, both also do not generate logical universality on their own, something we show pieceable fault-tolerance can indeed provide.

Achieving this universality also requires fundamentally new tools. For instance, we develop original circuit designs, called round-robin constructions, to perform our logical gates. Additionally, we create an adaptive procedure for error-correction on the non-stabilizer intermediate codes we encounter. The profitable use of non-stabilizer codes is perhaps novel and interesting in it of itself. A fault-tolerance overview, the definition of pieceable fault-tolerance, and a summary of these new tools are the goals of the next section.

\section{A design methodology for fault-tolerant logical gates}\label{design_method}
Quantum codes operate by the ``fight entanglement with entanglement" \cite{Preskill1998} mantra --- to protect sensitive data from a noisy, nosy environment, introduce additional degrees of freedom and encode the data in globally entangled states. Local errors have no chance of rearranging the long range entanglement to affect the data, and so our information is secure.

However, if we want to legitimately alter the data, this same security becomes a hassle. To perform quantum gates on the encoded data, we are forced to create circuits manipulating globally entangled states, and moreover, these circuits must not corrupt the data, even if they themselves are faulty! In this section, we overview this design challenge, culminating in our solution, pieceably fault-tolerant logical gates.

\subsection{Codes, logical operations, and error-correction}\label{dm_1}
An $\llbracket n,k\rrbracket$ quantum code $L$ using $n$ physical qubits to encode $k$ data qubits is essentially a collection of $2^k$ orthogonal $n$-qubit code states $\{\ket{\overline{0}},\ket{\overline{1}},\dots,\ket{\overline{2^k}}\}$. The code space $\mathcal{C}_L$ is the span of the code states. In this sense, quantum codes are nothing more than a strange basis for a subspace of a larger Hilbert space.

However, to be practically useful the code states must also satisfy certain error-correcting conditions \cite{Knill1997}. Not least of all, the code states should be far enough separated that errors localized to only a few physical qubits cannot undetectably change any code states. Also, it would be preferable if such error channels could be, not only detected, but reversed and the code space restored to normal.

Stabilizer codes \cite{Gottesman1996,Calderbank1997,Gottesman1997} were the first broad class of quantum codes that could be designed with such error-correcting properties, and they are still the most common, and promising, type of quantum code today. Stabilizer codes are built on the foundation of the Pauli group. The Pauli operators on a single-qubit are the familiar $\sigma_0=I$, $\sigma_1=X$, $\sigma_2=Y$, $\sigma_3=Z$. The Pauli group on $n$ qubits $\mathcal{P}_n$ is made of tensor products of these,
\begin{equation}
\mathcal{P}_n=\{i^a\bigotimes_{j=1}^n\sigma_{h_j}:a,h_j\in\{0,1,2,3\}\}.
\end{equation}
As for notation, given a Pauli operator $p\in\mathcal{P}_n$ we will denote the number of non-identity members of the tensor product as the weight $|p|$ and the set of qubits on which $p$ acts non-identically as the support $\text{supp}(p)$.

A subgroup of $\mathcal{P}_n$, called the stabilizer group $S$, defines the code space $\mathcal{C}_L$ of a stabilizer code as the $+1$ eigenspace of all Pauli operators in $S$. In order for this to be nontrivial, we must have $-I^{\otimes n}\not\in S$, which is equivalent to two conditions: first, $S$ is abelian, and second, all $g\in S$ have signs $\pm1$ (no sign of $\pm i$). Because every Pauli operator has only $\pm1$ eigenspaces of equal size, we evidently need $n-k$ independent Pauli operators to generate $S$ and reduce the $2^n$ dimensional Hilbert space of $n$ qubits to $2^k=2^n/2^{n-k}$ dimensional, the size of the code space. These generators are not unique, but nevertheless we will label (a canonical choice of) them $\widetilde Z_1,\widetilde Z_2,\dots\widetilde Z_{n-k}$ and write
\begin{equation}
S=\langle\widetilde Z_1,\widetilde Z_2,\dots,\widetilde Z_{n-k}\rangle,
\end{equation}
where $\langle\cdot\rangle$ indicates a list of generators rather than a list of all group elements.

Yet, this cannot be all there is to a stabilizer code, because only the code space, and not the code states, has so far been defined. To complete the stabilizer code, choose real-signed Pauli operators $\overline Z_1,\dots, \overline Z_k$ and $\overline X_1,\dots,\overline X_k$ from $\mathcal{P}_n$ that are independent from yet commute with both the stabilizer generators and each other. The only exception is for $\overline Z_i$ and $\overline X_i$, which should anticommute for each $i$. Since any set of $n$ independent Pauli operators from $\mathcal{P}_n$ have a unique state in their simultaneous $+1$ eigenspace, the full group $\langle\widetilde Z_1,\dots,\widetilde Z_{n-k},\overline Z_1,\dots,\overline Z_k\rangle$ defines a state $\ket{\overline{0}}$. The remaining encoded basis states are given by $\ket{\overline{b}}=\prod_j\overline X_j^{b_j}\ket{\overline{0}}$ for all $n$-bit strings $b$. Evidently, the $\overline X_j$ operators act as encoded, or logical, $X$ gates and the $\overline Z_j$ operators act as logical $Z$ gates. Logical $Y$ gates are given by $\overline Y_j=i\overline X_j\overline Z_j$.

However, there is some freedom in choice of logical operator even now. Indeed, any member of the stabilizer coset $\overline Z_i S$ acts exactly as $\overline Z_i$ does on the code space, and likewise with the cosets $\overline X_iS$ and $\overline Y_iS$ compared with $\overline X_i$ and $\overline Y_i$. The union of these $3k$ cosets and of the stabilizer itself forms the normalizer,
\begin{equation}
\mathcal{N}(S)=\langle\widetilde Z_1,\dots,\widetilde Z_{n-k},\overline Z_1,\dots,\overline Z_k,\overline X_1,\dots,\overline X_k\rangle.
\end{equation}
The lowest weight element of $\mathcal{N}(S)\setminus S$ has weight equal to the code distance $d$, which can be conveniently thought of as the fewest number of qubits which need to be acted upon to undetectably alter the code states. The double bracket notation for quantum codes is often extended from $\llbracket n,k\rrbracket$ to include the code distance as in $\llbracket n,k,d\rrbracket$.

Nevertheless, logical versions of the Pauli operators are not enough to perform arbitrary operations on encoded data. We should define logical versions of other unitary gates, say $U$, as well. Since the Pauli group $\mathcal{P}_k$ is a complete basis for the $k$-qubit unitaries, we can decompose $U=\sum_{p_j\in\mathcal{P}_k}a_jp_j$ for some $a_j\in\mathbb{R}$. A logical version of $U$, call it $\overline U$, should be similarly decomposed as $\overline U=\sum_{p_j\in\mathcal{P}_k}a_j\overline p_js_j$, where $s_j\in S$ and $\overline p_j$ is the logical version of $p_j$ wherein each occurrence of $X_i,Y_i,$ or $Z_i$ is replaced by $\overline X_i,\overline Y_i,$ or $\overline Z_i$, respectively. The freedom of the stabilizer is present in the form of the arbitrary $s_j$. Slightly more generally, we define a logical operation as any operator $\mathcal{U}$ that preserves the code space --- that is, $\mathcal{U}(\ket{\overline\psi})\in\mathcal{C}_L$ for all $\ket{\overline\psi}\in\mathcal{C}_L$.

Where does error-correction fit in this picture of code states and code spaces? Well, one thing an error-correction operation EC should not do is change code states. That is, $\text{EC}(\ket{\overline{\psi}})=\ket{\overline{\psi}}$ for all $\ket{\overline{\psi}}\in\mathcal{C}_L$. At the risk of over-generalization, we will actually take this condition as our definition --- an error-correction operation is any operation that preserves all code states. Note that even the trivial $I^{\otimes n}$ is an error-correction operation under this definition. We define error-correction this way because it makes our later definition of pieceable fault-tolerance most general.

The usual notion of error-correction we shall call complete error-correction $\overline{\text{EC}}$. In this case, for any Pauli error $E$ (note that all error channels can be decomposed into Pauli errors \cite{Nielsen2010}) of weight less than $d/2$, $\overline{\text{EC}}(E\ket{\psi})=\ket{\psi}$. There are several known methods for complete error-correction on quantum codes, including methods due to Shor \cite{Shor1996}, Steane \cite{Steane1997, Steane1998}, and Knill \cite{Knill2005}. We assume that every stabilizer code $L$ comes equipped with a canonical method of complete error-correction and denote it $\overline{\text{EC}}_L$.

\subsection{Circuits, faults, and fault-tolerance}\label{dm_2}
It is important to distinguish between what a logical operation or error-correction does, which we defined in the previous section as code space preserving or code state preserving, respectively, from its implementation, a circuit. Circuits are constructed from basic operations on physical qubits, including gates, $\ket{0}$ state preparation, and single-qubit measurement. The collection of allowed physical operations, or components, will be denoted $A_P$. A circuit $C$ then is nothing more than an ordered list of these physical operations $C=\{c_m,c_{m-1},\dots,c_1\}$. If the sequential action of these components, from right to left, performs a logical unitary gate, then the circuit can be called a logical circuit, and if it performs error-correction, then it can be called an error-correction circuit. To be most realistic, we could insist on dividing a circuit into several time steps and require each qubit to be acted upon during each time step, even if it is only by an identity operation. For most purposes we will treat circuits as operators and combine them with multiplication $C_2\cdot C_1$, with $C_1$ applied before $C_2$.

The fundamental problem we face in quantum error-correction is that the physical components of any circuit can be faulty and introduce errors to the qubits they act upon. The standard model of faulty components, used in essentially every discussion of fault-tolerance \cite{Knill1996, Zalka1996, Gottesman1997, Aliferis2005}, restricts the way that a component $c\in A_P$ can fail by saying that any faulty version of $c$ can be modeled as $c$ followed by an arbitrary error channel acting on the qubits in the support of $c$. A correlated error between different qubits therefore only arises when those qubits are explicitly coupled by a physical component. It is an interesting and important fact that these analog-like error channels can be ``digitized" \cite{Nielsen2010}. That is, from the perspective of an error-correcting code, any error channel can be viewed as the random application of an (unintended) Pauli operator. In turn, these Pauli operators can be decomposed into single-qubit Pauli errors. For example, we already discussed that any complete error correction of a distance $d$ quantum code can correct $t<d/2$ single-qubit errors.

The solution to the problem of faulty components is to design fault-tolerant circuits. Such circuits are created with a quantum code $L$ with a complete error-correction operator $\overline{\text{EC}}_L$ in mind. We say a logical circuit $C$ is fault-tolerant (with respect to $L$) if any single faulty component in the combined circuit $C\cdot\overline{\text{EC}}_L$ creates only errors correctable by an ideal version of $\overline{\text{EC}}_L$ performed afterward. Whether the circuit is fault-tolerant or not might be said to be judged by this final ideal $\overline{\text{EC}}_L$. Since errors arising from faulty components in the middle of $C\cdot\overline{\text{EC}}_L$ must be propagated to the end of the circuit before being corrected, it is crucial that components of $C$ do not couple too many qubits within a single code block so that the spread of errors is limited. Also, in correspondence with the exREC formalism \cite{Aliferis2005}, it is important to include the leading complete error-correction, because input errors to $C$ arise from faults in the previous round of error-correction.

The definitions of fault-tolerance and of faulty components suggest the following sufficient design methodology for logical gates --- a logical circuit is evidently fault-tolerant if all qubits interact (either directly or indirectly) with fewer than $t$ qubits in any other code block and fewer than $t-1$ qubits in their own blocks and $t<d/2$. In this case, called $t$-transversality \cite{Jochym2014}, the standard model of faulty components implies that any one code block cannot accumulate more than $t$ single-qubit errors assuming only a single faulty component. For $d=3$ codes, the essential concept is $1$-transversality, a property which most current logical circuit designs possess.

However, it can be proved that transversality is not enough --- for no quantum code can we create a quantum universal set of logical gates with transversal circuits alone \cite{Eastin2009, Zeng2011}. This fact has prompted other techniques of logical circuit design to complete a universal logical gate set. Some of the largest breakthroughs have been universality with transversal Clifford gates and magic ancillas \cite{Bravyi2005}, with a transversal gate set but with a gauge qubit reset \cite{Paetznick2013}, with non-transversal constructions on large codes built as concatenations of smaller codes with complementary transversal gate sets \cite{Jochym2014}, and with code-switching techniques \cite{Hill2013,Anderson2014} to switch between codes with complementary transversal gate sets.

\subsection{Pieceable fault-tolerance}\label{dm_3}
We now define our method to provide universality without transversality. We are motivated by a simple idea: even if an entire logical circuit is not fault-tolerant, parts of the circuit may be. If we perform error-correction partway through the circuit, perhaps we can quell the propagation of errors before they become uncorrectable. This idea was used before in the fault-tolerant code switching technique of Hill et.~al.~\cite{Hill2013}, and here we provide a generalization, one that succeeds in making even some non-Clifford circuits fault-tolerant.

To make this idea precise, begin with a logical circuit $C$ for some logical gate for a quantum code $L$. Decompose this logical circuit into pieces $C=C_m\cdot C_{m-1}\cdot\dotsc\cdot C_1$. After each partial logical gate $C_i$, there is still a code space and code states, and if $L$ was a stabilizer code there is still a stabilizer and logical Pauli operators (though each may be non-Pauli if $C_i$ is non-Clifford). Thus, we can insert code state preserving operations, error-corrections $\text{EC}_i$, after each piece without altering the logical effect of $C$. We now have a modified circuit,
\begin{equation}\label{mr_ft}
\widetilde C=\text{EC}_m\cdot C_m\cdot\text{EC}_{m-1}\cdot C_{m-1}\cdot\dotsc\cdot\text{EC}_1\cdot C_1.
\end{equation}
If such a modified circuit can be found and $\widetilde C$ is fault-tolerant in the traditional sense, then we say that $C$ is pieceably fault-tolerant (in $m$ pieces). We also make a distinction between any ancilla qubits used during the circuit pieces $C_i$, called functional ancillas, and those used during $\text{EC}_i$, called error-correction ancillas. All of our constructions use no functional ancillas, although pieceable constructions with more pieces will obviously use more error-correction ancillas than a non-pieceable (i.e. $m=1$) logical gate.

A few things should be clarified about the definition. First, a traditionally fault-tolerant circuit $C$ is evidently pieceably fault-tolerant in one piece because $\widetilde C=\text{EC}_1\cdot C=C$ is traditionally fault-tolerant when we take $\text{EC}_1$ to be the trivial error-correction $I^{\otimes n}$. 

Second, we include the final error-correction $\text{EC}_m$ even though the code space has already returned to that of $L$ because, for full generality, we may want to perform a round of error-correction before we have our circuit's fault-tolerance judged by the ideal $\overline{\text{EC}}_L$, as the exREC definition says. For instance, we will see examples where the intermediate error-corrections share classical syndrome data with $\text{EC}_m$ so that the set of errors $\text{EC}_m$ corrects is different than that $\overline{\text{EC}}_L$ would. If $\text{EC}_m$ is not needed, however, it can always be set to $I^{\otimes n}$ and remain in line with the definition. 

Third, whenever the stabilizer becomes non-Pauli, intermediate error-corrections will be trickier than simply CAT state measurement of all stabilizers. In those cases, we must be extra careful to construct circuits that correct all errors we need to correct without introducing errors that cannot be corrected in the next round. The quantum error-correcting conditions \cite{Nielsen2010}, although providing a necessary condition for such error-correcting circuitry to exist, do not provide a sufficient one.

Finally, note that we do not require an intermediate error-correction $\text{EC}_j$ to correct all errors resulting from one fault in the previous circuitry. Although sufficient to guarantee pieceable fault-tolerance, such a condition is not necessary. Indeed, to simplify the circuitry of $\text{EC}_j$ and mitigate the number of errors it might introduce itself, we will often intentionally let errors slip by an intermediate error-correction as long as they are correctable in some subsequent round (often the final round).

At this point, it may not actually be clear that pieceable fault-tolerance is going to do anything interesting. Nevertheless, our spirits might be buoyed by the example in Hill et.~al.~\cite{Hill2013}. After all, being able to switch from the 5-qubit to 7-qubit codes in a pieceably fault-tolerant fashion immediately endows the 5-qubit code with a pieceably fault-tolerant CZ gate, because the 7-qubit code has a transversal one. 

In Section~\ref{5qubit}, we will construct a significantly simpler CZ gate for the 5-qubit code by a pieceably fault-tolerant construction in two pieces. Our CZ construction also suggest a similar design for a pieceably fault-tolerant logical CCZ gate on the 5-qubit code. Notably, this completes a universal set of gates for the 5-qubit code, which already has a transversal set of single-qubit Cliffords. In Section~\ref{7qubit}, we develop a similar CCZ construction for the 7-qubit code. Moreover, in Section~\ref{conditions_for_pieceably}, we show that all nondegenerate stabilizer codes have a pieceably fault-tolerant logical gate equivalent to CCZ under local Clifford gates. Actually, there is more that is possible with pieceable fault-tolerance. For any two \emph{different} nondegenerate stabilizer codes, we can perform some logical gate locally Clifford equivalent to a CZ gate between them. This for instance allows pieceably fault-tolerant SWAP gates between different codes (when they have appropriate fault-tolerant local Cliffords), providing perhaps the broadest method of code switching yet known, at least when no functional ancillas are used. Finally, each of these statements generalizes to $\text{C}^h\text{Z}$ and $\text{C}^h\text{X}$ gates, where $h$ is an integer specifying the number of control qubits (e.g. $\text{C}^1\text{Z}$ is a CZ gate and $\text{C}^2\text{Z}$ is a CCZ. We will denote what is usually called CNOT by CX and Toffoli by $\text{C}^2\text{X}=\text{CCX}$).

\subsection{Additional pieceable concepts}\label{dm_4}
We take the time now to introduce some terminology applying to pieceably fault-tolerant circuits in general. We use these concepts, too, for our specific constructions.

Let us first try to envision what the code stabilizer might look like after several pieces $C_{1r}=C_r\cdot C_{r-1}\cdot\dotsc\cdot C_1$ of the logical circuit. We will assume $C_{1r}$ is unitary for this discussion. Basic stabilizer theory tells us that, if the stabilizer group began as $S\le\mathcal{P}_n$, then after $C_{1r}$, we will have $S_r=\{C_{1r}sC_{1r}^\dag:s\in S\}$. If $C_{1r}$ is non-Clifford, then at least some members of $S_r$ may be non-Pauli. How does error-correction work for such a stabilizer?

\begin{figure}
\includegraphics[width=\columnwidth]{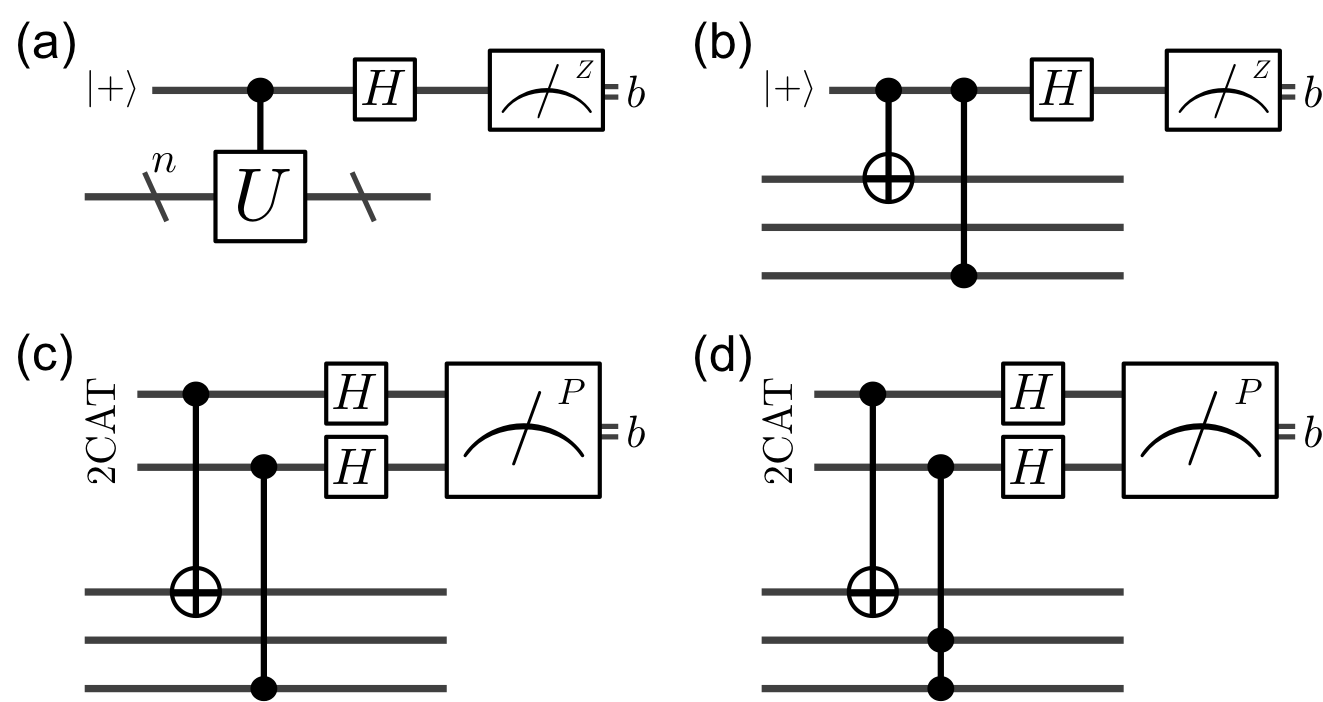}
\caption{\label{fig0} (a) The method to measure a hermitian and unitary $n$-qubit operator $U$ \cite{Nielsen2010}. When the result of the $Z$-basis measurement is $b=\pm1$ the state of the lower qubit is projected into the $\pm1$ eigenspace of $U$. (b) The method applied to measuring a Pauli $XIZ$ on three code qubits. (c) The method made fault-tolerant by using a verified 2-qubit CAT state, $2\text{CAT}=(|00\rangle+|11\rangle)/\sqrt{2}$ \cite{Shor1996}. The measurement is now of the parity of the ancilla qubits. (d) The CAT state method for measuring a non-Pauli operator $X_1\text{CZ}(2,3)$. Strictly speaking this measurement is not fault-tolerant, because the failure of the CCZ gate can cause two errors on the code qubits (which is generally uncorrectable if the code has distance $3$). This is why we are so careful about measuring non-Pauli operators.}
\end{figure}

First, note that while members of $S_r$ might not be Pauli, each is at least both hermitian and unitary, because all $s\in S$ were. Each also still has two eigenspaces, with eigenvalues $\pm1$, that are equally sized. The circuit in Fig.~\ref{fig0}a still serves to measure elements of $S_r$. So, still, we can measure a generating set of $S_r$. However, as already mentioned in Section~\ref{dm_3}, we must be careful about what errors a failure in these measurements might introduce to the next piece of the circuit.

So we first ask, what form can these errors take? We first need to recognize that some errors are contagious, in the sense that they propagate other errors within some piece $C_i$. We define the set of contagious errors of a pieceable circuit $C$ as
\begin{equation}
\mathcal{E}_C=\{E\in\mathcal{P}_n:\exists i\text{ s.t. }[E,C_i]\neq0\}.
\end{equation}
All other errors, however, those that commute with all pieces $C_i$, are errors that may be allowed to pass an intermediate error-correction (or be introduced by faults in one). Also, we can never prevent error-correction from introducing single-qubit errors to the next piece, and so single-qubit errors must also be allowable errors.

So our goal at intermediate error-correction should be to correct all contagious errors and do so in such a way that only noncontagious and single-qubit errors are ever possibly allowed to enter the next piece of the circuit. An immensely helpful tool in this regard is the constant stabilizer of the circuit $C$, which we define as
\begin{equation}\label{gen_const_stab}
S_C=\{s\in S:\forall i,[s,C_i]=0\}.
\end{equation}
The benefit of $S_C$, which note is a subgroup of the whole stabilizer $S$, is that all its members remain Pauli throughout the entirety of the pieceable circuit $C$. The upshot is that we can always measure elements of $S_C$ with the CAT state method, Fig.~\ref{fig0}c, and be assured that this process will not introduce more than a single-qubit error to the code qubits. The other stabilizers, those we call nonconstant because they do not commute with some piece of $C$, are not guaranteed to have the same property, failing specifically when they are non-Pauli.

This suggests an adaptive error-correction procedure to be used whenever we have both constant and non-Pauli nonconstant stabilizers. We will require that constant stabilizers can detect all contagious errors. The procedure begins by measuring the constant stabilizers. If an error is not detected, then we know that no contagious errors have occurred. We can therefore let whatever errors have occurred (if any) pass through uncorrected. If an error has been detected, then we know a fault has occurred, and, barring a second fault, future stabilizer measurements will be faultless. Thus, we can now measure the nonconstant stabilizers, deduce what error has occurred, and correct it.

Notice that the constant stabilizers have been used to guarantee that the measurement of the nonconstant stabilizers proceeds reliably, without any faults. We will call this a reliability guarantee on the nonconstant stabilizers. This is necessary because faulty measurement of nonconstant stabilizers can introduce contagious errors to the next piece of the circuit, which may allow errors to spread out-of-hand. Consider for instance failure of the CCZ gate in Fig.~\ref{fig0}d.

A second guarantee can be given for a nonconstant stabilizer as well, which, while not strictly necessary, is convenient when we want to think about errors as Paulis. To define this guarantee, imagine that we have a set of Pauli errors $\mathcal{E}\subseteq\mathcal{P}_n$ entering an intermediate error-correction stage. Rather than applying a non-Pauli recovery, as the generalized error-correction conditions described above say to do, we want to apply Pauli recovery operations only. Measurement of a stabilizer generator $g$ eliminates the possibility of some Pauli errors, but to measure $g$ requires knowing that each possible Pauli error remaining either commutes or anticommutes with $g$. For generators of the constant stabilizer $g\in S_C$ this is trivial, but for a nonconstant stabilizer $t\in S\setminus S_C$, there is no reason a priori to assume the commutation guarantee for $t$ is true. 

The way we solve this problem is with a four-step procedure --- (1) measure the constant stabilizers, (2) partially correct some Pauli errors so that commutation guarantees exist on nonconstant stabilizers, (3) measure the nonconstant stabilizers, and (4) complete the recovery. Notice that steps (1), (3), (4) we had already planned to use to assure a reliability guarantee. Step (2) has been added so that recovery, though now split into two stages (2) and (4), consists entirely of Pauli operators. We call this our Procedure for Adaptive, Reliable Stabilizer Error-Correction or PARSEC, and it will play a prominent role in our non-Clifford pieceably fault-tolerant constructions. 

We note that PARSEC generally has these four-steps, but because it is adaptive some may be skipped in any one instance, and there are other variations depending on the code. The PARSEC variants we find most useful in general will be detailed later as Algorithms \ref{parsec} and \ref{cssparsec}.

As for pieceable constructions, one type of circuit is our main innovation in logical circuits and makes our systematic study of pieceable fault-tolerance possible. We call it the round-robin design. Given an $h$-qubit gate $U$ and disjoint sets of qubits $\Lambda_1,\Lambda_2,\dots,\Lambda_h$, the round-robin circuit of $U$ on $\{\Lambda_i\}$ is
\begin{equation}\label{round_robin_def}
\prod_{j_1\in\Lambda_1}\prod_{j_2\in\Lambda_2}\dots\prod_{j_h\in\Lambda_h}U(j_1,j_2,\dots,j_h).
\end{equation}
We also make this definition assuming that $U$ is such that that all gates in the product mutually commute, so that ordering the products is unnecessary. Then, all of our new pieceably fault-tolerant logical constructions are in fact of the round-robin variety with different choices of $U$ and the sets $\Lambda_i$. When it comes to round-robin circuits, we will refer to qubits in $\Lambda=\bigcup_{j=1}^h\Lambda_j$ as active qubits and any non-ancilla qubits not in $\Lambda$ as idle.

\section{The 5-qubit code}\label{5qubit}

Having overviewed the design of logical circuits in theory and defined our particular method, we now get our hands dirty and do some actual circuit designing, beginning with the 5-qubit code. We show how to make fault-tolerant CZ and CCZ gates on the 5-qubit code. More than these are possible in the formalism of pieceable fault-tolerance, but we save the most general proofs until Section~\ref{conditions_for_pieceably}. Our goal here is to present concrete examples of pieceably fault-tolerant logic on a small quantum code. This includes examples of contagious errors, constant stabilizers, PARSEC, and round-robin circuits, concepts which were introduced only very broadly in the previous section.

\subsection{Transversal gates}

The 5-qubit quantum code, a $\llbracket5,1,3\rrbracket$ stabilizer code, was one of the earliest quantum codes to be discovered \cite{Bennett1996,Laflamme1996} and is the smallest quantum code of distance $3$. The stabilizer and normalizer of the 5-qubit code can be written as
\begin{align}
S_5&=\left\langle\begin{array}{c}ZZXIX\\XZZXI\\IXZZX\\XIXZZ\end{array}\right\rangle\\
\overline Z_5&=-XIZIX\\
\overline X_5&=-YIXIY.
\end{align}
Notice that we have deviated from the standard presentation of $\overline Z_5=ZZZZZ$ and $\overline X_5=XXXXX$ to versions that are equivalent under multiplication by stabilizers, and also have the lowest possible weight. 

\begin{figure}
\includegraphics[width=\columnwidth]{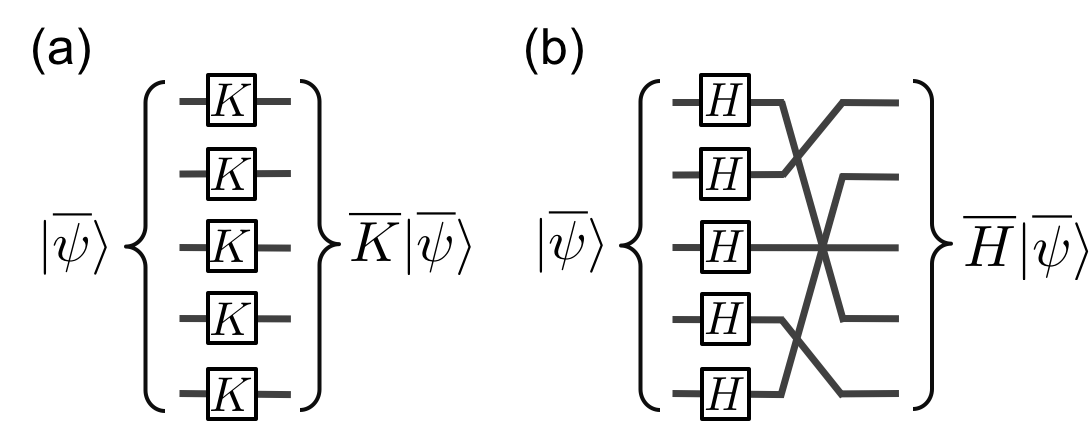}
\caption{\label{fig1} Transversal implementations of (a) the octahedral gate $K=SH$ and (b) the Hadamard gate $H$ on the 5-qubit code.}
\end{figure}

It is well-known that the 5-qubit code supports a universal, transversal set of single-qubit Clifford operations. To begin with, for $s_x,s_y,s_z\in\{-,+\}$, all eight members of the family of single-qubit octahedral gates
\begin{equation}\label{octahedral_gate}
K_{s_xs_ys_z}=e^{i\frac{\pi}{3\sqrt{3}}\left(s_xX+s_yY+s_zZ\right)},
\end{equation}
are transversal on the 5-qubit code in the obvious way $\overline K_{s_xs_ys_z}=K_{s_xs_ys_z}^{\otimes 5}$. A special case that we will name and use often is $K_{+++}=SH:=K$, where $S=\text{diag}(1,i)$ and $H=(X+Z)/\sqrt2$. The set of transversal single-qubit Clifford operations can therefore be completed by implementing a logical Hadamard gate as $\overline H=P_\pi H^{\otimes 5}$, where $P_\pi$ is an appropriate permutation of the five qubits, one example of which is shown in Fig.~\ref{fig1}b. A permutation is transversal if it is implemented, not by physical SWAP gates, but by simply relabeling the qubit lines. It is known, at least for stabilizer codes, that such permutations alone cannot bypass the universality-transversality no-go \cite{Zeng2011}.

Gottesman \cite{Gottesman1998} gave a way to complete the set of fault-tolerant Clifford operations on the 5-qubit code. He noted that a three-qubit Clifford gate $T_3$ is transversal on the 5-qubit code, and supplemented with measurement and single-qubit Clifford gates, can provide a fault-tolerant implementation of CX as well. There are a couple downsides to this approach. First, $T_3$ is a complicated gate. To implement $T_3$ on physical qubits will likely require compiling into simple one- and two-qubit Clifford gates, as experiments will not have direct access to $T_3$. Second, implementing logical CX with $T_3$ requires the use of five functional ancilla qubits making up the third 5-qubit code block that is eventually measured and discarded. 

\subsection{Pieceably fault-tolerant CZ}\label{5cz}
Our first goal is to develop a pieceably fault-tolerant logical circuit for CZ on the 5-qubit code that uses no functional ancilla qubits. In fact, our pieceably fault-tolerant circuit consists of only two pieces. Moreover, we will be able to easily generalize this design to pieceably fault-tolerant logical CCZ (in four pieces) on the 5-qubit code.

\begin{figure}
\includegraphics[width=\columnwidth]{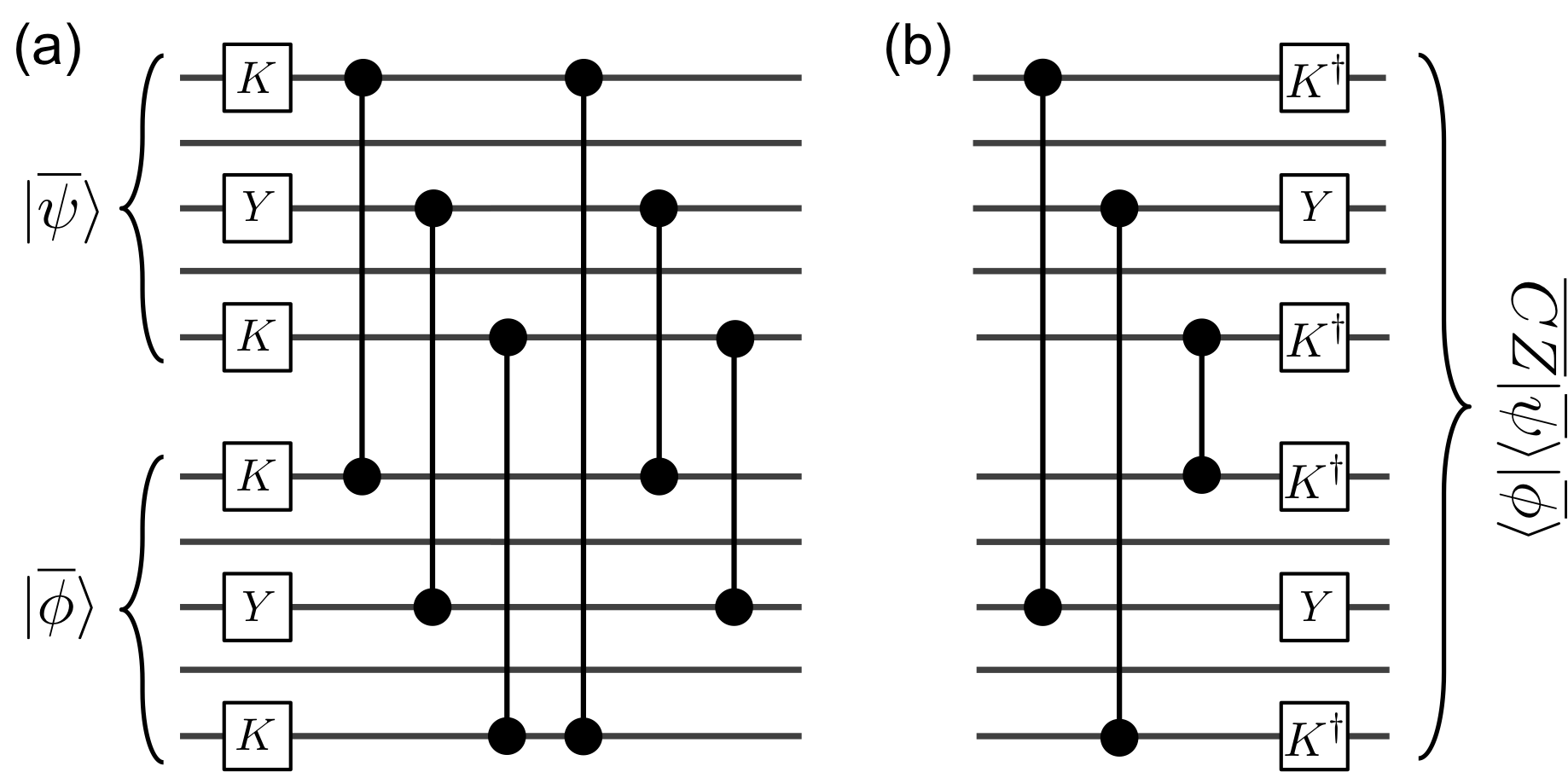}
\caption{\label{fig2} Circuits $C_a$ (a) and $C_b$ (b) such that ${C_{\text{CZ}}=C_b\cdot C_a}$ is a pieceably fault-tolerant implementation of CZ on the 5-qubit code. The intermediate error correction, done after (a), is described in the text, while the final error correction after (b) can be taken as two copies of the canonical 5-qubit error-correction, which is the canonical error-correction for the tensor product of two 5-qubit codes. In the notation of the definition of pieceable fault-tolerance ${\widetilde C_{\text{CZ}}=C_b\cdot\text{EC}_1\cdot C_a}$. To do a logical CX from the top code block to the bottom change all physical CZs to physical CXs from qubits in the top to those in the bottom. The intermediate error-correction will still succeed in correcting all single faults if all stabilizers of the corresponding intermediate code are measured.}
\end{figure}

Both pieces of the pieceably fault-tolerant circuit for logical CZ are shown in Fig.~\ref{fig2}. This is our first example of a useful round-robin construction, accompanied by a prologue and epilogue of local Clifford gates. The story begins with a local unitary transformation $K_1Y_3K_5$ on both code blocks. These Clifford gates take the 5-qubit code to an equivalent $\llbracket5,1,3\rrbracket$ code, which we will refer to as the $5'$-qubit code, with
\begin{align}\label{sprime}
S_5'&=\left\langle\begin{array}{c}-YZXIZ\\-ZZZXI\\-IXZZZ\\-ZIXZY\end{array}\right\rangle\\
\overline Z_5'&=ZIZIZ\\
\overline X_5'&=XIXIX.
\end{align}
Label qubits in the first code block $j_A$ and qubits in the second code block $k_B$. Now it is simple to check that the round-robin circuit of CZ gates,
\begin{equation}\label{cz_combo}
\prod_{j,k\in\{1,3,5\}}\text{CZ}(j_A,k_B),
\end{equation}
which contains nine CZ gates total, preserves the combined stabilizer of both blocks $\{s_1\otimes s_2:s_1,s_2\in S_5'\}$. This is most easily seen by noticing that in each stabilizer generator there are an even number of $X$ and $Y$s on the active qubits, qubits $1,3,5$ of each block. This round-robin circuit also effects the appropriate transformation on the normalizers of codeblocks $A$ and $B$,
\begin{alignat}{2}\label{CZ_action}
\overline Z_A\otimes\overline I_B&\rightarrow\overline Z_A\otimes\overline I_B,\text{\space\space} \overline I_A\otimes\overline Z_B&&\rightarrow\overline I_A\otimes\overline Z_B,\\
\overline X_A\otimes\overline I_B&\rightarrow\overline X_A\otimes\overline Z_B,\text{\space\space} \overline I_A\otimes\overline X_B&&\rightarrow\overline Z_A\otimes\overline X_B,
\end{alignat}
due to the odd number of $X$s on active qubits in $\bar X_5'$.

Breaking the round-robin circuit Eq.~\eqref{cz_combo} after any six CZs that touch no single qubit more than twice, is sufficient to guarantee pieceable fault-tolerance. One example of such a decomposition is shown in Fig.~\ref{fig2}. To use the language of Section~\ref{dm_2}, the pieces should be 2-transversal. The intuitive explanation of why this works, is that, while the 5-qubit code can only correct single-qubit errors when errors are randomly distributed across all code qubits, if errors are restricted to two known locations, then all errors on those two qubits are correctable. By sharing information about the stabilizers they have measured, code blocks $A$ and $B$ can both correct individual, uncorrelated errors if they exist, and inform each other about possible propagation of errors from one block to the other, a propagation that will affect at most two known qubits in the other block.

Anthropomorphizing aside, we can actually check that all errors introduced by components in Fig.~\ref{fig2}(a) are correctable by measuring the stabilizers of the appropriate code after Fig.~\ref{fig2}(a). The code in question is a $\llbracket10,2,3\rrbracket$ stabilizer code with stabilizer generators
\begin{equation}\label{cz_intermediate_s}
S_{10}=\left\langle\begin{array}{cc}
-YZXIZIIZIZ&-ZIZIIZIXZY\\
-ZZZXIIIIII&-IIIIIIXZZZ\\
-IXZZZIIIII&-IIIIIZZZXI\\
-ZIXZYZIIIZ&-ZIIIZYZXIZ
\end{array}\right\rangle.
\end{equation}
Note that the first and last rows of generators, those with Pauli weight on both blocks, are nonconstant stabilizers, and the remaining four generators are constant stabilizers, since they commute with the round-robin circuit, Eq.~\eqref{cz_combo}. We will also refer to the generators on the left as belonging to block $A$ and those on the right as belonging to $B$, since they are the transformations of the generators in Eq.~\eqref{sprime} for their respective blocks.

Assume only one component in Fig.~\ref{fig2}a has failed, and propagate the resulting Pauli errors to the end of the subcircuit, where we will measure the stabilizer generators in Eq.~\eqref{cz_intermediate_s} in a fault-tolerant fashion. A generator is said to trigger if we measure that stabilizer and find a $-1$ eigenstate.

There are three cases from the measurement: (1) constant stabilizers trigger in both blocks (2) constant stabilizers from just one block trigger (3) no constant stabilizers trigger. The same three cases will appear in later arguments as well. They are convenient since the constant stabilizers of a block \emph{not} triggering implies no contagious errors in that block, which proves that the nonconstant stabilizers of the \emph{other} block have triggered (or not) depending only on errors within their native block.

We proceed to argue that errors within case (1) can be corrected. Here, we know that a CZ gate has failed with contagious errors ($X$ or $Y$s) on both nodes. These may have also propagated $Z$ errors through a subsequent CZ. Fortunately, the constant stabilizers, when restricted to the active qubits, represent a bit-flip redundancy code $\langle ZZI,IZZ\rangle$ on each block, so we can pinpoint exactly the qubit lines with contagious errors, diagnose precisely which CZ failed, and learn where $Z$ errors have propagated. At most two errors are present in each block, with known locations. Now notice that we can obtain the eigenvalues of any original $5'$-qubit stabilizer (i.e. Eq.~\eqref{sprime}) on either block using the measured syndromes, the eigenvalues of the Paulis in Eq.~\eqref{cz_intermediate_s}, and our knowledge of the locations of contagious errors. Thus, the error-correction problem now becomes two independent corrections of at most two located errors per block of the distance three, $5'$-qubit code.

In case (2), if the constant stabilizers of, say, block $A$ triggered, we know that there is no contagious error in block $B$, and moreover only a single error in block $A$. Thus, the syndrome of all block $A$'s stabilizers, both constant and nonconstant, will determine the error in $A$ in the usual 5'-qubit code fashion for single error-correction. If that error was contagious, we will also know now the two possible locations of $Z$ errors in block $B$. As before in case (1), by calculating the eigenvalues of the original $5'$-qubit code generators for block $B$ from our measured syndromes plus knowledge of the error in block $A$, we can correct the two errors in the usual $5'$-qubit code fashion for correcting two located errors.

Finally, case (3) implies that no contagious errors have occurred, and also that there is at most a single-qubit error in each block. Therefore, the syndrome will independently correct errors in both blocks in usual 5'-qubit code fashion.

It is interesting to consider that the intermediate error-correction procedure we have just described actually uses only 84 non-trivial syndromes out of a total $2^8-1=255$ possible. The remaining syndromes do not correspond to errors that can arise in the standard model of faulty components from just one component failing, but they will allow the correction of some errors arising from two faulty components.

It should be noted that the final error-correction for our logical CZ construction can be taken to be a canonical error-correction, consisting of two copies of the canonical error-correction for the 5-qubit code. This works because the final piece of the round-robin circuit, Fig.~\ref{fig2}b, is 1-transversal.

\subsection{$\overline T$ will always require functional ancillas}\label{5t}
We have described how to obtain the entirety of the Clifford group fault-tolerantly on the 5-qubit code, without any ancillas other than those needed for error-correction. A final logical gate is needed, however, to complete a universal set of fault-tolerant gates. From the perspective of universality, there are many choices we could make --- any gate outside the Clifford group will do \cite{Zeng2011}. Nevertheless, from the perspective of fault-tolerance, some gates are indeed forbidden from being implemented without overhead in the form of functional ancillas. The $T=\text{diag}(1,e^{i\pi/4})$ gate is one of these. 

\begin{figure}
\includegraphics[width=\columnwidth]{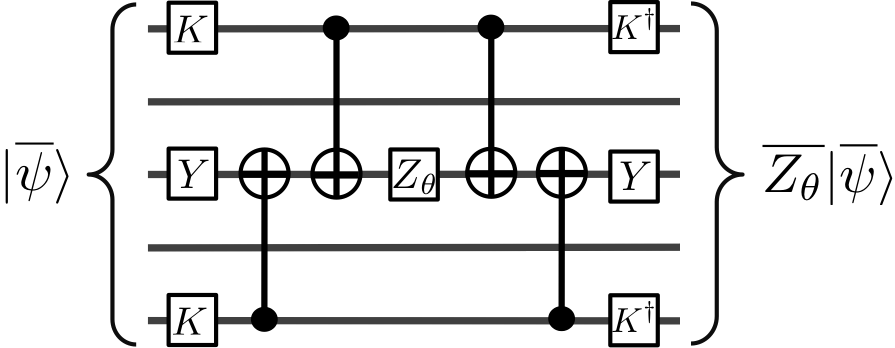}
\caption{\label{fig3} A logical $Z_\theta=\exp(i\theta Z/2)$ gate for the 5-qubit code. The case $\theta=\pi/4$ corresponds to a $T$-gate. This construct, like any logical $T$-gate circuit using no functional ancillas, is not fault-tolerant nor even pieceably fault-tolerant. However, this does not mean such designs are entirely useless. For instance, the above circuit provides a fault-tolerant circuit for $T$ in the $\llbracket75,1,3\rrbracket$ code formed by concatenating the $\llbracket15,1,3\rrbracket$ Reed-Muller code \cite{Knill1996, Steane1996c} within the $5'$-qubit code (the code resulting from applying $K_1Y_3K_5$ to the 5-qubit code, see Eq.~\eqref{sprime}), because the Reed-Muller code has both transversal $T$ and transversal CX logical constructs. This same idea of concatenating codes to produce a larger code with a greater, and sometimes even universal, set of fault-tolerant logical gates is used to great effect in \cite{Jochym2014}.}
\end{figure}

To show that a fault-tolerant $T$-gate on the 5-qubit code cannot be implemented without functional ancillas, even in a pieceably fault-tolerant fashion, begin by noting that a logical $T$ cannot be implemented with only single-qubit gates and qubit swaps --- that is, $T$ is not a transversal gate on the 5-qubit code. This follows from the second theorem of Zeng et. al. \cite{Zeng2011}, where they show that no stabilizer code has a universal, transversal set of gates even on one encoded qubit. Since $T$ is not transversal, any logical $T$ construction with no ancillas must use a multi-qubit gate between some number of the five code qubits (e.g. see Fig.~\ref{fig3}). Look at the chronologically last such multi-qubit gate. After that gate, if the circuit really implements a logical $T$ gate, the codespace will be locally unitary equivalent to the codespace of the 5-qubit code. Thus, because the 5-qubit code is a perfect code, we know that no multi-qubit errors will be correctable (as long as all single-qubit errors must simultaneously be correctable, which is always the case). We conclude that the last multi-qubit gate will always have the potential to introduce uncorrectable errors if we use no functional ancillas to implement $T$ on the 5-qubit code. 

The same argument works for any non-transversal single-qubit gate and any code that is unable to correct all errors on all subsets of multiple code qubits. For instance, the 7-qubit code also needs functional ancillas for a logical $T$-gate, because the classical codes underlying its CSS construction are perfect.

When functional ancilla qubits are used, a pieceably fault-tolerant logical circuit for $T$ does exist for the 5-qubit code by our construction of a pieceably fault-tolerant CCZ in the next section and the universality that it bestows. But it is also interesting to note other constructions in the literature. First, there are the magic state constructions due to Shor \cite{Shor1996} and Bravyi and Kitaev \cite{Bravyi2005}. Alternatively, code switching techniques, first from the 5-qubit to 7-qubit code \cite{Hill2013}, and then from the 7-qubit to 15-qubit code \cite{Paetznick2013}, also give a fault-tolerant method of implementing $T$, since it becomes transversal once we have transferred to the 15-qubit code.

\subsection{Pieceably fault-tolerant CCZ}\label{5ccz}
In this section, we describe a pieceably fault-tolerant implementation of the CCZ gate on the 5-qubit code, thereby completing a fault-tolerant universal set of gates for the smallest quantum error-correcting code. Unlike fault-tolerant $T$-gate implementations, this construction requires no functional ancilla qubits. It is however a pieceably fault-tolerant design of four pieces, so there are four rounds of error-correction to contend with. The error-correction procedures themselves are novel, of the adaptive variety termed PARSEC in Section~\ref{dm_4}, since, unlike the logical CZ case of Section~\ref{5cz}, intermediate in the logical CCZ circuit the code space is no longer that of a stabilizer code. 

\begin{figure*}
\includegraphics[width=\textwidth]{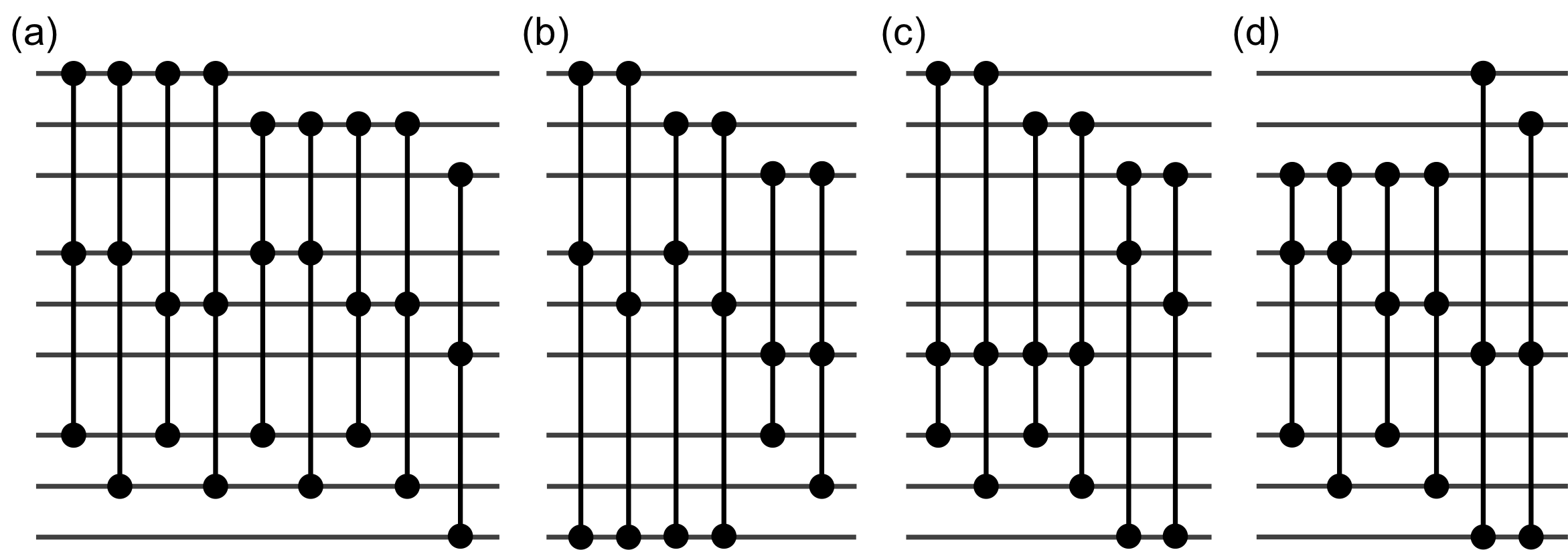}
\caption{\label{fig4} The round-robin circuit for CCZ between three code blocks each with three active qubits (e.g. qubits $1,3,5$ for a 5'-qubit code block and qubits $5,6,7$ for a 7-qubit code block). Idle code qubits are omitted. The circuit is broken into four pieces (a), (b), (c), (d), each of which is 2-transversal, and therefore an appropriate error-correction performed after each piece will ensure pieceable fault-tolerance (see text). This circuit design is quite versatile. For instance, this idea performs a CCZ on the $5'$-qubit code (Section~\ref{5ccz}), on the 7-qubit code (Section~\ref{7ccz}), and between $5'$- and 7-qubit code blocks, among other possibilities (Section~\ref{conditions_for_pieceably}). On some codes (see Section~\ref{conditions_for_pieceably} for explicit conditions), including the $5'$- and $7$-qubit codes, replacing the CCZs with CCXs or CCYs will also implement logical CCX or logical CCY, respectively.}
\end{figure*}

The pieceably fault-tolerant CCZ circuit is shown in Fig.~\ref{fig4}. The general idea is no different from the design of the fault-tolerant CZ. First, transform all three code blocks into the 5'-qubit code in Eq.~\eqref{sprime} by applying the single-qubit unitaries $K_1Y_3K_5$. Second, apply the round-robin CCZ circuit
\begin{equation}\label{ccz_combo}
\prod_{j,k,l\in\{1,3,5\}}\text{CCZ}(j_A,k_B,l_C),
\end{equation}
containing $3^3=27$ CCZs total. Finally, transform each block back to the standard 5-qubit code with $K_1^\dag Y_3K_5^\dag$.

Checking that Eq.~\eqref{ccz_combo} implements a logical CCZ on the 5'-qubit code is only slightly more difficult than checking that Eq.~\eqref{cz_combo} implements logical CZ. The $\text{CCZ}=\text{diag}(1,1,1,1,1,1,1,-1)$ gate is in the third level of the Clifford hierarchy \cite{Gottesman1999}. Moreover, it treats all three qubits symmetrically, commutes with Pauli $Z$s, and acts on Pauli $X$s like
\begin{equation}\label{CCZ_action}
\text{CCZ}(XII)\text{CCZ}=XII\times\text{CZ}(2,3).
\end{equation}
We must verify the same action of Eq.~\eqref{ccz_combo} on the 5'-qubit codes' logical operators $\overline Z_A,\overline Z_B,\overline Z_C,\overline X_A,\overline X_B,\overline X_C$ and preservation of the stabilizer.

The latter is simple. Since each stabilizer generator has an even number of $X$ or $Y$ operators on the active qubits (again qubits $1,3,5$ of each block), any $\text{CZ}$s introduced to the stabilizer by Eq.~\eqref{CCZ_action} will cancel by the end of the round-robin circuit, Eq.~\eqref{ccz_combo}.

As for the transformation of the logical operators, we can note that $\overline Z_Q$ for code blocks $Q=A,B,C$ trivially does not change. Since $\overline X_Q$ is odd weight and is supported by Pauli $X$s on only the active qubits, we find
\begin{equation}\label{logical_CCZ_action}
\overline X_A\overline I_B\overline I_C\longrightarrow \overline X_A\overline I_B\overline I_C\prod_{j,k\in\{1,3,5\}}\text{CZ}(j_B,k_C),
\end{equation}
and likewise for $\overline X_B$ and $\overline X_C$. However, we have already argued in Section~\ref{5cz} that the latter product of CZs, the round-robin CZ circuit from Eq.~\eqref{cz_combo}, is exactly a logical CZ on the 5'-qubit code (in this case on blocks $B$ and $C$), and so Eq.~\eqref{logical_CCZ_action} is the logical version of Eq.~\eqref{CCZ_action}, as required.

We now show that the round-robin CCZ circuit of Eq.~\eqref{ccz_combo} is pieceably fault-tolerant. The decomposition proceeds similarly to that of the round-robin CZ circuit. Group the CCZs of Eq.~\eqref{ccz_combo} into pieces such that, in any piece, no single qubit interacts with more than two qubits of another block (i.e. each piece is 2-transversal). One possibility, with four pieces, is shown in Fig.~\ref{fig4}. We will perform error-correction after each piece, to be described shortly. 

However, let us first note the form of the stabilizer at any of the three intermediate error-corrections. There are four stabilizer generators corresponding to each block, for a total of twelve. For instance, the four from block $A$ at any intermediate error-correction will have the form,
\begin{equation}\label{ccz_intermediate_s}
S_A=\left\langle\begin{array}{l}-YZXIZ\times\text{CZ}_{A1}\\-ZZZXI\\-IXZZZ\\-ZIXZY\times\text{CZ}_{A4}\end{array}\right\rangle,
\end{equation}
where it is understood that these generators contain $I^{\otimes5}$ on blocks $B$ and $C$ and $\text{CZ}_{A1}$ and $\text{CZ}_{A4}$ are products of CZs between blocks $B$ and $C$. For instance, after the first piece of Fig.~\ref{fig4}, $\text{CZ}_{A4}=\text{CZ}(1_B,1_C)\text{CZ}(1_B,3_C)\text{CZ}(3_B,1_C)\text{CZ}(3_B,3_C)\text{CZ}(5_B,5_C)$. The stabilizer generators for blocks $B$ and $C$ are of a similar form. Again, two stabilizer generators per block, six total, have changed (the nonconstant stabilizers) and two per block have not (the constant stabilizers). 

Of course, the nonconstant stabilizers have not changed when we just look at the qubits on their native block. We will use this fact, because if we have ensured at most $Z$ errors in the other blocks (this is the meaning of commutation guarantees in this case), then the nonconstant stabilizers in block $A$ (for instance) will act just as the original stabilizer generators in terms of detecting errors on $A$.

Assume that a single faulty component in the previous circuitry, say $\text{EC}_{k}\cdot C_k\cdot\dots\cdot\text{EC}_1\cdot C_1$, will, if anything, introduce either noncontagious errors or single-qubit errors to the $(k+1)^{\text{th}}$ piece, $C_{k+1}$. It can be verified that the intermediate error-correction procedure we describe next actually does ensure this --- only noncontagious errors will pass, and, assuming no earlier fault in the circuit, a failure in the error-correction itself will introduce at most a single-qubit error. The latter property is because we only measure nonconstant stabilizers when we have a reliability guarantee (see Section~\ref{dm_4}).

\begin{algorithm}[t]
	\caption{\label{parsec} PARSEC}
    \SetKwInOut{Input}{Input}
    \SetKwInOut{Output}{Output}
    
    \Input{(1) $h+1$ code blocks partway through a round-robin circuit.\newline
    (2) Each block has an error-correcting constant stabilizer.}
    \Output{(1) $h+1$ code blocks free of contagious errors.\newline
    (2) The locations of noncontagious errors if there are multiple per block.}
    
    $\bullet$ Measure the constant stabilizers of all blocks\\
    $\bullet$ \textbf{If} two or more blocks have triggered: (\emph{faulty $\text{C}^h\text{Z}$}) \\
    \Indentp{2em}
    	$-$ \textbf{For} each triggered block: \\
    \Indentp{2em}
    		- Apply $X$ to the affected qubit \\
    		- Note possible spread of $Z$ errors to other blocks \\
    \Indentp{-4em}
    $\bullet$ \textbf{If} one block triggered: (\emph{single error in that block}) \\
    \Indentp{2em}
    	$-$ Measure nonconstant stabilizers of that block \\
        $-$ Correct the error \\
        $-$ Note possible spread of $Z$ errors to other blocks \\
    \Indentp{-2em}
    $\bullet$ \textbf{If} no blocks triggered: (\emph{at most noncontagious errors}) \\
    \Indentp{2em} 
    	$-$ Do nothing \\
    \Indentp{-2em}
    \hrulefill\\
A procedure for doing intermediate error-correction in the non-stabilizer codes encountered partway through the round-robin construction of logical CCZ and, in Section~\ref{conditions_for_pieceably}, the constructions of Theorem~\ref{thm1}. An ``error-correcting constant stabilizer'' is one that can distinguish between all contagious errors modulo noncontagious errors, defined more formally in Section~\ref{conditions_for_pieceably}.
\end{algorithm}

We implement error-correction $\text{EC}_{k+1}$ using PARSEC, Alg.~\ref{parsec}. But why does it work for the 5-qubit code? We go through the procedure step-by-step. Begin by measuring the constant stabilizers of each block (six total). Since these are Paulis, we can measure them in standard fashion (e.g. via a CAT state ancilla \'{a} la Shor \cite{Shor1996} with repeats and majority voting). Notice that any contagious errors will be detected by these constant stabilizer measurements. Indeed if constant stabilizers from two or three different blocks have triggered (case (1)), we know for sure that a CCZ has failed, since this is the only way just one failure could introduce errors detectable by the constant stabilizers into two different blocks. We also know that the errors are contagious. These errors are immediately correctable modulo noncontagious errors --- we know where the errors have occurred from the constant stabilizers' syndrome, and can apply $X$ to the afflicted qubits, leaving at most $Z$ errors. If constant stabilizers have just triggered in one block (case (2)), then we know that at worst noncontagious errors are present in the other blocks, and that the triggered block has at most a single-qubit error. If no constant stabilizers have triggered (case (3)), then at worst noncontagious errors have occurred.

In case (1), having found and corrected contagious errors, we know where $Z$ errors might possibly be. By the way the pieces are constructed (i.e. 2-transversally), at most two $Z$ errors will be present in each block. Now we could just measure all nonconstant stabilizers (see for example Fig.~\ref{fig0}d), with reliability and commutation guarantees from having detected and corrected the $X$ errors, and this would give enough information to correct the $Z$ errors. But luckily, the $Z$ errors will not propagate further, so we can also just leave them until the final error-correction, when all stabilizers are Pauli again and so simple to measure. This means we have to inform the final error-correction of the locations of these $Z$ errors, and this is why that information is listed as an ``output" in Alg.~\ref{parsec}. If the final error-correction gets no information from the intermediates, it should revert to canonical form, correcting one arbitrary error per block.

In case (2), if the constant stabilizers of block $Q$ have triggered we have, assuming just one faulty component, reliability and commutation guarantees on the nonconstant stabilizers of block $Q$. Measuring them (and this time, but only this time, the nonconstant stabilizers must be measured), we can deduce what and where the error in block $Q$ is. Correct the error in block $Q$. If the error had occurred on an idle qubit, we are done. If the error was on an active qubit, however, then we know where the possible $Z$ errors may be in other blocks. Again, inform the final error-correction.

Though case (3) might appear the most straightforward, it is actually the most interesting. In this case, we could go ahead and measure all nonconstant stabilizers of all blocks, and therefore find the locations of, at most, one $Z$ error per block. However, there is a problem with this approach --- we have no reliability guarantee! Having detected no error, the measurement of the nonconstant stabilizers during $\text{EC}_{k+1}$ could itself contain the first fault of the circuit, and it might result in a contagious error entering the next piece of the circuit $C_{k+2}$. This is a problem, because to analyze of cases (1) and (2) we assumed one faulty component in $\text{EC}_{k}\cdot C_k\cdot\dots\cdot\text{EC}_1\cdot C_1$ causes either a noncontagious error or a single-qubit error to enter $C_{k+1}$, and the same assumption must hold for $k\rightarrow k+1$. We therefore cannot measure the nonconstant stabilizers in case (3).

The solution to this problem is, of course, that we do not actually need to correct the $Z$ errors, as we have already seen. They will not propagate further, and so can just be corrected at the final error-correction. Since there is only at most one per block, we do not even need to give the final error-correction special instruction.

Because the CCZ and CZ gates we have designed for the 5-qubit code can recover from only one faulty component, we envision concatenating \cite{Aliferis2005} the circuits $k$ times to achieve tolerance to at most $2^k-1$ faults. Relevant to this process, the CCZ circuit we designed in this section has the appealing property of requiring only physical components (be it gates, measurements, or state preparations) that can be directly implemented logically on the 5-qubit code. Therefore, concatenating the circuit of Fig.~\ref{fig4} can be done without first compiling the CCZ gates into 1- and 2-qubit gates. Only at the physical level, and only in architectures without native CCZ gates, must such a compilation be done.

Next, we mention the how the same PARSEC procedure is general enough to work on the 7-qubit code as well. But it can also be made simpler, by virtue of the 7-qubit code being CSS.

\section{7-qubit code}\label{7qubit}
\subsection{Transversal gates}
The 7-qubit code \cite{Steane1996} is the most frequently studied of the small stabilizer codes, trumping the 5-qubit code in this metric despite its larger size. One advantage of the 7-qubit code is its complete set of transversal Clifford gates --- all three of $H$, $S$, and CNOT are transversal in the simplest way, as 7 copies of the physical gate applied qubitwise (e.g. $\overline H=H^{\otimes7}$). It is also a code grounded in classical coding theory, being the smallest member of the CSS code family.

Here we briefly recall the stabilizer and normalizer of this $\llbracket7,1,3\rrbracket$ code. Normalizers, like those of the 5-qubit code previously, are reduced to minimum weight.
\begin{align}\label{s7}
S_7&=\bigg\langle\begin{array}{cc}XXXXIII&ZZZZIII\\XXIIXXI&ZZIIZZI\\XIXIXIX&ZIZIZIZ\end{array}\bigg\rangle\\
\overline Z_7&=IIIIZZZ\\
\overline X_7&=IIIIXXX.
\end{align}
Transversality of $H,S,$ and CNOT can easily be verified.

\subsection{Pieceably fault-tolerant CCZ}\label{7ccz}
In this section, we complete a universal set of gates on the 7-qubit code. Initially, we present a round-robin design mirroring that for the 5-qubit code and fitting within our general framework for pieceable gate design in Section~\ref{conditions_for_pieceably}. However, the 7-qubit code has other properties that lead to appreciable optimization, which we discuss and use to perform a resource comparison against magic state injection in the following section.

A logical CCZ circuit for the 7-qubit code circuit is the round-robin CCZ circuit
\begin{equation}\label{ccz_combo_7}
\prod_{j,k,l\in\{5,6,7\}}\text{CCZ}(j_A,k_B,l_C)
\end{equation}
with error-correction after each subcircuit of CCZs. These circuit pieces, like those in the case of the 5-qubit code, should be made 2-transversal. Qubits $5,6,7$ of each block are active qubits and the remainder are idle qubits.

Checking that Eq.~\eqref{ccz_combo_7} implements a logical CCZ follows the same logic as in the 5-qubit case. The stabilizer generators will all be preserved because the first three each have an even number of $X$s on the active qubits and the last three commute with Eq.~\eqref{ccz_combo_7}. The normalizer $\overline Z_7$ will also be preserved for the same commutation reason. However, $\overline X_7$ will change,
\begin{equation}\label{ccz_xbar_transform}
\overline X_A\overline I_B\overline I_C\longrightarrow\overline X_A\overline I_B\overline I_C\prod_{j,k\in\{5,6,7\}}\text{CZ}(j_B,k_C)
\end{equation}
with the analogous transformation on the normalizers of blocks $B$ and $C$. The thing to note is that the round-robin product of CZs in Eq.~\eqref{ccz_xbar_transform} is also, just like for the 5-qubit code, a logical CZ on the 7-qubit code. And so Eq.~\eqref{ccz_xbar_transform} is the correct logical transformation consistent with a logical CCZ.

The pieces of Eq.~\eqref{ccz_combo_7} consist of one group of nine CCZs and three groups of six, just like the 5-qubit case (see Fig.~\ref{fig4} again). The 7-qubit code also has some stabilizer generators that become non-Pauli as the round-robin CCZ circuit progresses, namely the second and third, the nonconstant stabilizers, and some generators that do not change at all, the first and the last three. Although the first is a constant stabilizer, it is trivially so (all Paulis on active qubits are $I$s), so only the last three, those of $Z$-type, need be concerned when we say to measure constant stabilizers. Our job now is to argue that the pieces in Fig.~\ref{fig4} can all be error-corrected without introducing uncorrectable errors into the next piece.

The key to intermediate error-correction in the 7-qubit case is that all contagious $X$ errors can be corrected by measuring only the constant stabilizers, while noncontagious $Z$ errors commute through the circuit to the end. Because the 7-qubit code is CSS it can detect and correct one $X$ error per block using only measurement of the first three stabilizers in Eq.~\eqref{s7}. Conveniently and not coincidentally, at most one $X$ or $Y$ error will occur per block as well, resulting from failure of a physical CCZ. Having learned the locations of $X$ or $Y$ errors, the intermediate error-correction can then tell the final error-correction where to look for $Z$ errors, at most two of them per block in known locations. The final error-correction measures all stabilizers (which are Pauli once we have returned to the tensor product of three 7-qubit code spaces) and applies recovery using its syndrome plus the information from any intermediate error-corrections.

\begin{algorithm}[t]
	\caption{\label{cssparsec} CSS PARSEC}
    \SetKwInOut{Input}{Input}
    \SetKwInOut{Output}{Output}
    
    \Input{(1) $h+1$ code blocks partway through a round-robin circuit.\newline
    (2) Each block has an error-correcting constant stabilizer.\newline
    (3) Each block's constant stabilizer also distinguishes between contagious errors and errors on idle qubits.}
    \Output{(1) $h+1$ code blocks free of contagious errors.\newline
    (2) The locations of noncontagious errors if there are multiple per block.}
    
    $\bullet$ Measure the constant stabilizers of all blocks\\
    $\bullet$ \textbf{If} any block has triggered: \\
    \Indentp{2em}
    	$-$ \textbf{For} each triggered block: \\
    \Indentp{2em}
    		- If error is contagious, apply $X$ to affected qubit and note spread of $Z$ errors to other blocks \\
    		- If error is on an idle qubit, do nothing \\
    \Indentp{-4em}
    $\bullet$ \textbf{If} no blocks triggered: \\
    \Indentp{2em} 
    	$-$ Do nothing \\
    \Indentp{-2em}
    \hrulefill\\
An error-correction procedure achieving the same goals as PARSEC, but without ever having to measure the nonconstant stabilizers of any block. The CSS in the procedure's name stems from the fact that any normalizer of a CSS code that consists entirely of $X$s and $I$s or of $Z$s and $I$s will have a constant stabilizer satisfying the third input condition (but not necessarily the second, though nondegeneracy of the code is sufficient to do so).
\end{algorithm}

Notice, no non-Pauli stabilizers were ever measured! Yet, we still managed to correct any single component failure in the circuit Fig.~\ref{fig4}. Not having to measure nonconstant stabilizers is a feature of all our round-robin constructions on CSS codes. Indeed, this modified PARSEC we call CSS PARSEC, outlined in Alg.~\ref{cssparsec}.

\subsection{A resource comparison against magic states}\label{resource_compare}
The most popular method for performing logical non-Clifford gates on the 7-qubit code is through magic state injection \cite{Bravyi2005,Nielsen2010}. While many other codes require an extensive distillation procedure to create magic states, the 7-qubit code offers a postselective strategy \cite{Aliferis2005, Goto2016}, which is less resource intensive. So it is the latter which we compare against in this section, using three metrics: ancillas consumed, multi-qubit gate counts, and circuit volume (timesteps times active qubits). All the circuits used for this comparison are drawn in Appendix D.

Specifically, the postselection protocol we compare against is a highly optimized version similar to that of Goto \cite{Goto2016} (whose circuit produces $H$-type magic states for implementing $T$ gates). With our circuit, CCZ-states for implementing a teleported CCZ gate are produced also after just one measurement of the appropriate non-Pauli operators and one syndrome measurement. This circuit is less resource intensive than an AGP-type postselection \cite{Aliferis2005} that measures non-Pauli operators multiple times and also less intensive than decomposing a logical CCZ into four $T$ gates \cite{Jones2013} that are each magically injected (e.g. using \cite{Goto2016}). For ease of resource counting, we assume that any postselections always succeed so our metrics do not have functional dependence on any physical error-rates. This assumption generally helps the magic state method more, because of the large circuit volume that must be recomputed upon failed postselection.

As for the pieceable construction, while the circuit in the previous section for logical CCZ uses 27 physical CCZs, there is actually a simpler design that uses 21 physical CCZs and needs just one intermediate error-correction. We will briefly describe the two main ideas of this optimization. The first idea is to exploit the fact that CZ is transversal on the 7-qubit code to show that
\begin{equation}\label{ccz_combo_7b}
\prod_{j,k\in\{5,6,7\}}\text{CCZ}(j_A,k_B,k_C)
\end{equation}
implements logical CCZ. Unfortunately, as written Eq.~\eqref{ccz_combo_7b} needs to be broken into four pieces to be made pieceably fault-tolerant. This can be traced to the fact that qubits $5,6,7$ from block $A$ are each coupled to seven qubits from the other blocks. In accordance with the fault-tolerance principle of keeping pieces at most 2-transversal, we see $\lceil7/2\rceil=4$ pieces are required for pieceable fault-tolerance.

To reduce the number of pieces, the second idea is to use the $Z$-stabilizers of block $A$ to spread the nodes of the 21 CCZs around to all seven of the code qubits of block $A$. Indeed, we can do so such that each code qubit in any block is involved in no more than three CCZ gates. Thus, 2-transversality implies just $\lceil3/2\rceil=2$ pieces are required. See Fig.~\ref{shortest_ccz} in Appendix D for one particular fault-tolerant piecing.

We use the two piece, 21 CCZ pieceably fault-tolerant circuit derived above in our resource comparison. Error-correction, both intermediate and final, are done with Steane states created by postselection that is assumed to always succeed.

Table~\ref{tab1} shows the final resource comparisons. We find that using a pieceable CCZ gate over magic state injection nearly \emph{halves} the required resources for logical CCZ. These results also give a reason to expect a lower logical error-rate for a pieceable CCZ, though we leave the exact determination of this quantity to future work.

\begin{table}
\begin{tabular}{|c|c|c|c|c|}
\hline
 & CX-count & CCZ-count & ancillas & volume \\\hline
Magic states & 312 & 22 & 132 & 1518 \\
Pieceable & 162 & 21 & 72 & 771 \\\hline
\% improvement & 48.1\% & 4.5\% & 45.5\% & 49.2\% \\\hline
\end{tabular}
\caption{\label{tab1} Resource comparison for implementing $\overline{\text{CCZ}}$ on the 7-qubit code using either magic state injection via postselection or pieceable fault-tolerance.}
\end{table}

\section{Conditions for pieceable fault-tolerance in stabilizer codes}\label{conditions_for_pieceably}
The previous two sections, Section~\ref{5qubit} about constructing pieceably fault-tolerant circuits for the 5-qubit code, and Section~\ref{7qubit} about similar constructions for the 7-qubit code, left one burning theoretical question open: for what codes do there exists pieceably fault-tolerant constructions for interesting gates? 

In this section, we explore this question in detail. we find several remarkable results, including the fact that all nondegenerate stabilizer codes with distance at least three have a pieceably fault-tolerant gate that is locally Clifford equivalent to $\text{C}^h\text{Z}$. For codes already endowed with transversal local Clifford gates, this is enough for universality. Our general constructions also work between blocks in different codes as well, implying a method of code switching with pieceable circuits.

Because we found round-robin circuits so useful in Sections~\ref{5qubit} and \ref{7qubit}, we will continue studying them in this section, only in more generality. In the first part of this section, we have two main concerns. Our first is to make logical gates from round-robin circuits. Only after do we state further conditions so that the logical circuits are also pieceably fault-tolerant, which comes alongside a general method for piecing round-robin circuits into few pieces. In the second part of this section, our concern is the identification of well-known codes and families of codes that can attain universal sets of logical gates through pieceable fault-tolerance.

\subsection{Pieceable constructions in general}
To find logical uses for round-robin circuits, we define a particular $h+1$ qubit gate, a generalization of $\text{C}^h\text{Z}$. Given a trit string $\{j_0,j_1,\dots,j_h\}$ with $j_m\in\{1,2,3\}$, let
\begin{equation}
\Gamma(\sigma_{j_0},\sigma_{j_1},\dots,\sigma_{j_h})=A^\dag\left(\text{C}^h\text{Z}\right)A,
\end{equation}
where $A=\bigotimes_{m=0}^hK^{3-j_m}$. Recall that $(\sigma_1,\sigma_2,\sigma_3)=(X,Y,Z)$ and $K=SH$. Notice that $\Gamma(Z,Z,\dots,Z)=\text{C}^h\text{Z}$. Likewise, $\Gamma(X,Z,\dots,Z)$ is a  $\text{C}^h\text{X}$ with target the $0^\text{th}$ qubit, and $\Gamma(X,X,\dots,X)$ is a symmetric gate analogous to $\text{C}^h\text{Z}$ but in the $X$-basis. We will also abuse notation slightly to write $\Gamma$-gates in terms of Pauli operators, like $\Gamma(p)$ where $p\in\mathcal{P}_{h+1}$ and $|p|=h+1$. So $\Gamma(Z^{\otimes h+1})=\text{C}^h\text{Z}$ as well.

With round-robin circuits we can at least implement logical versions of all $\Gamma$-gates on all stabilizer codes. We save the argument for fault-tolerance of the construction for later.
\begin{thm}\label{thm1}
Consider any $h+1$ stabilizer codes $L_0,L_1,L_2,\dots,L_h$ each encoding a single qubit. For all $h$ and all $p\in\mathcal{P}_{h+1}$ with $|p|=h+1$, there is a round-robin circuit, perhaps conjugated by local Clifford gates, that implements logical $\Gamma(p)$.
\end{thm}
\emph{Proof.} Choose a logical $p_j$ operator for code $j$ and denote it $\overline{p}_j$. For instance, if $p_j=Z$ then $\overline{p}_j$ indicates a logical $Z$ operator of code $j$. With these choices of logical operators made, build the following circuit.
\begin{enumerate}
\item By applying local Clifford gates, change every $\overline p_j$ into a form consisting of only Pauli $Z$s and $I$s with positive sign. This of course changes the stabilizer and other normalizers of the codes accordingly. We call this the $Z$-form of the code with respect to $\overline{p}_j$.
\item Apply the round-robin circuit of $\text{C}^h\text{Z}$ on the sets $\text{supp}(\overline p_0),\text{supp}(\overline p_1),\dots,\text{supp}(\overline p_h)$.
\item Reverse the Clifford gates of step (1) to return the codes to their original form.
\end{enumerate}
The only non-trivial part of proving this ``prologue, round-robin, epilogue'' 
construction works is in showing that a round-robin $\text{C}^h\text{Z}$ circuit acts as logical $\Gamma(p)$ on codes in $Z$-form. To that purpose, from now on we will just assume the $h+1$ codes are already in $Z$-form.

How does $\Gamma(p)$ act on Paulis? It should be clear, generalizing Eq.~\eqref{CZ_action} for the CZ and Eq.~\eqref{CCZ_action} for the CCZ, that $\Gamma(p)q_j\Gamma(p)$ is $q_j$ if $q_j\in\{I,p_j\}$ and $q_j\Gamma(p_0,\dots,p_{j-1},p_{j+1},\dots,p_h)$ otherwise (where it is to be understood that the $h$-qubit $\Gamma$-gate here applies to all qubits except qubit $j$). We have to argue the same statements hold logically after the round-robin circuit of $\text{C}^h\text{Z}$s and, moreover, that the stabilizer is preserved.

Tackle the latter point first. A stabilizer of the $j^{\text{th}}$ code $s\in S_j$, since it commutes with $\overline{p}_j$ and $\overline{p}_j$ is in a form consisting of all $Z$s or $I$s, must have an even number of $X$s and $Y$s on $\text{supp}(\overline{p}_j)$. Thus, the round-robin $CZ$ circuit will preserve all stabilizers $s$. Likewise, $\overline{p}_j$ is also preserved.

Now, to argue the other normalizers transform correctly, we argue by induction on $h$. Start with $h=1$, where the round-robin circuit is of CZ gates. A normalizer $l\in\mathcal{N}(S_0)\setminus S_0$ of the $0^\text{th}$ code that is not a logical $p_0$ operator, will necessarily anticommute with $\overline{p}_0$. Therefore, $l$ has an odd number of $X$s and $Y$s on $\text{supp}(p_0)$ and so $l\rightarrow l\otimes p_1$ under the action of the round-robin CZ circuit. The symmetric argument holds for the $1^{\text{st}}$ code block.

Now the inductive step with $h+2$ code blocks. The same fact stands that for code $j$ if $l\in\mathcal{N}(S_j)\setminus S_j$ and $l\not\in \overline{p}_jS_j$, then $\{l,\overline{p}_j\}=0$ and so $l$ has an odd number of $X$s and $Y$s on $\text{supp}(\overline{p}_j)$. However, now this means that
\begin{align*}
l\rightarrow l\prod_{\substack{q_i\in\text{supp}(\overline p_i)\\i\neq j}}\text{C}^h\text{Z}(q_0,\dots,q_{j-1},q_{j+1},\dots,q_{h+1}).
\end{align*}
But by induction, the product of $\text{C}^h\text{Z}$ gates is logical $\Gamma(p_0,\dots,p_{j-1},p_{j+1},\dots,p_{h+1})$. This is exactly the effect we wanted.
\hfill$\blacksquare$

To show that the circuits presented in the proof of Theorem~\ref{thm1} are pieceably fault-tolerant requires specification of where and how intermediate error-correction should be done. We will put additional conditions on a stabilizer code to guarantee that such intermediate error-correction can be done. The idea behind these conditions is simply a generalization of the procedure used in the 5-qubit and 7-qubit cases. At an intermediate error-correction, we will first measure the constant stabilizers, which remain Pauli throughout the round-robin circuit, and correct errors that we can correct before measuring any non-Pauli stabilizers (using some measurement circuit like Fig.~\ref{fig0}d). Our conditions amount to the existence of an appropriate group of constant stabilizers for {\it each} code involved.

Because each code can be different, it makes sense to place conditions on the individual codes rather than all $h+1$ codes together. To do this, we should define constant stabilizers for the individual codes. We could consider taking from the stabilizer $S$ of code $j$ only the stabilizers that commute with the round-robin circuit $C$. But recall from the proof of Theorem~\ref{thm1} that the round-robin circuit itself depends on a normalizer $p\in\mathcal{N}(S)\setminus S$ of code $j$ --- it is wired to all qubits in the support of $p$. Therefore, in the end, the constant stabilizer of code $j$ depends only on which normalizer $p$ we choose. We therefore propose the constant stabilizer {\it of a normalizer} be defined as
\begin{equation}\label{const_stab}
S_C(p)=\{g\in S:\forall i\in\text{supp}(p), [g_i,p_i]=0\}.
\end{equation}
Notice that $S_C(p)$ is a subgroup of $S$, and in fact $S_C(p)$ is a proper subgroup if and only if $d\ge2$, where $d$ is the code distance. Examples for the 5-qubit code include $S_C(ZZZZZ)=\{IIIII\}$ and $S_C(XIZIX)=\langle IXZZX,XZZXI\rangle$. We emphasize that this definition of the constant stabilizer by way of a normalizer, Eq.~\eqref{const_stab}, is really a special case of the more general definition, Eq.~\eqref{gen_const_stab}, when the pieceable circuit in question is a round-robin circuit like in Theorem~\ref{thm1}.

The same logic that changed the constant stabilizer from depending on the round-robin circuit to depending on a normalizer applies to the contagious errors. Accordingly, define the contagious errors of a normalizer $p$ (on an $n$-qubit code block) to be
\begin{equation}\label{cont_errs}
\mathcal{E}_C(p)=\{E\in\mathcal{P}_n:|E|=1,[E,p]\neq0\}.
\end{equation}
We have included the additional condition $|E|=1$, because after a single fault contagious errors can never affect more than one qubit per code block in our round-robin constructions.

It is important to study how $S_C(p)$ and $\mathcal{E}_C(p)$ interact. After all, we saw in the 5-qubit and 7-qubit code examples that the constant stabilizer detecting contagious errors is key to PARSEC. So we will say that $S_C(p)$ is error-detecting if all errors in $\mathcal{E}_C(p)$ anticommute with some member of $S_C(p)$. Likewise, we say $S_C(p)$ is error-correcting if it can distinguish between all contagious errors, at least modulo noncontagious errors. These definitions can be rephrased in terms of the distance of a certain classical code related to the stabilizer code and normalizer in question (see Appendix A). This relation to classical coding should not be unexpected --- after the local Clifford transformation of the codes to $Z$-form in Theorem~\ref{thm1}, the stabilizers with only $Z$s on the active qubits are the constant ones, and the contagious errors are $X$s (and $Y$s, but these are just $X$s modulo a noncontagious $Z$).

The following theorem establishes pieceable fault-tolerance for the round-robin constructions of Theorem~\ref{thm1}, along an upper bound on the number of pieces required.
\begin{thm}\label{thm3}
Consider $h+1$ stabilizer codes $L_0,L_1,L_2,\dots,L_h$ such that
\begin{enumerate}
\item Each $L_j$ encodes a single qubit and has code distance $d_j\ge3$. 
\item For all $j$, $L_j$ has a logical $p_j$ operator with an error-correcting constant stabilizer and weight, say, $w_j$. 
\end{enumerate}
Then, there is a pieceably fault-tolerant circuit for logical $\Gamma(p)$ using $\prod_{j=0}^hm_j/\min_j m_j$ pieces, where $m_j=\lceil w_j/(d_j-1)\rceil$.
\end{thm}\noindent
\emph{Proof.} 
We first establish that a decomposition $C=C_m\cdot C_{m-1}\cdot\dotsc\cdot C_1$ of the round-robin circuit $C$ is good, in the sense that intermediate error-correction after each $C_k$ will succeed in correcting all contagious errors, if for all $k=1,2,\dots, m$ and all $i,j\in\{0,1,\dots,h\}$ all qubits in $\Lambda_i$ are connected to at most $d_j-1$ qubits in $\Lambda_j$ by gates in $C_k$. Once we come up with an intermediate error-correction procedure that works given this connectivity condition, we will set about piecing the circuit to achieve it.

All intermediate error-corrections proceed according to PARSEC, Algorithm \ref{parsec}. We go through this general procedure now, and indicate where we use the assumptions of the theorem. First, measure all constant stabilizers of all the code blocks. There are now three cases to consider. 
\begin{enumerate}
\item If two or more blocks have triggered, we know that a $\text{C}^h\text{Z}$ has failed, because there is no other way to get a contagious error in more than one block. Since the constant stabilizers are error-correcting, and we know the errors are indeed contagious, we can locate the affected qubits and apply $X$, converting them to at most noncontagious $Z$ errors on the active qubits. However, we must also note which qubits the affected qubits had interacted with during the last piece of the circuit --- those qubits might have picked up $Z$ errors. Luckily, there are at most $d_j-1$ such qubits in block $j$ by our piecing of the circuit. Make note of those locations and inform the final error-correction.
\item If only one block (say $j$) has triggered, we do not yet know whether the constant stabilizers have detected a contagious error on the active qubits or an error on the idle qubits, but we do know there is at most a single-qubit error in this block. To find out if it is contagious, measure the nonconstant stabilizers of block $j$. Fortunately, we can do so (with a circuit like Fig.~\ref{fig0}(d)) because we have ensured a reliability guarantee on this nonconstant stabilizer measurement by having detected that some fault had already occurred. We also have commutation guarantees for the nonconstant stabilizers of block $j$, since no contagious errors have been detected on the other blocks. The nonconstant stabilizers of block $j$ have the form of a Pauli on the qubits of block $j$ and $\text{C}^{h-1}\text{Z}$ gates on the other blocks. The commutation guarantees assure us that only the Pauli part will be involved in error-detection. Since this Pauli part is the same as it was in the original code $L_j$, we have reduced the problem to correcting a single error with code $L_j$. Code $L_j$ can do this correction modulo its own stabilizers. However, the nonconstant stabilizers have since changed, so we must ensure that the error-correction does not happen modulo the nonconstant stabilizers. The fact that the constant stabilizer of block $j$ is error-correcting ensures this, however. More precisely, there can be no nonconstant stabilizers of code $L_j$ with weight two (see Appendix A, Lemma~\ref{lem1}(4)). After correcting the error, if it was contagious, make note of where it could have spread to other code blocks, and inform the final error-correction.
\item If no blocks have triggered, at most there is a single-qubit noncontagious error per block. These errors will not spread, and they can be corrected by the final error-correction.
\end{enumerate}
The final error-correction at the end of the round-robin circuit is special in a different way than the intermediate error-corrections. Since the codespace is once again stabilizer, we can measure a complete generating set. Normally, this would allow us to correct at least one error per block (since $d_j\ge3$ for all $j$), and it will do so if case (3) occurred in the intermediate error-corrections. However, if case (1) or (2) occurred in an intermediate error-correction then we will instead switch to correcting at most $d_j-1$ noncontagious errors in block $j$, using the information about the locations of those errors, which was sent to the final error-correction by the intermediate error-correction that detected them.

One might wonder if something like PARSEC is really necessary for intermediate error-correction? In Appendix B, however, we show that, provided we use lowest weight normalizers as our $\overline{p}_j$, intermediate in the round-robin circuit the codespace is non-stabilizer. Therefore, \emph{some} procedure different from the conventional stabilizer error-correction is required.

Now we can group the physical $\text{C}^h\text{Z}$ gates of the round-robin circuit $C$ according to the connectivity condition --- in any one piece of the circuit, a qubit is connected to at most $d_j-1$ qubits of block $j$. First, take the $w_j$ qubits of $L_j$ and partition them into sets of size at most $d_j-1$. Thus, $\Lambda_j=\lambda_{j1}\cup\lambda_{j2}\cup\dots\cup\lambda_{jm_j}$ where $m_j=\lceil w_j/(d_j-1)\rceil$. Note that $C=R_M\cdot R_{M-1}\cdot\dotsc\cdot R_1$ where $R_j$ are round-robin circuits of $\text{C}^h\text{Z}$ on sets $\lambda_{0i_0},\lambda_{1i_1},\dots\lambda_{hi_h}$ for some $i_k$ indices dependent on $j$ and $M=m_0m_1\dots m_h$. Effectively, each code block is now made of $m_j$ ``composite qubits" and $C$ can be viewed as a round-robin circuit of ``composite gates" (themselves round-robin circuits) on these composite qubits. The connectivity rule for pieces on the composite objects is that within a piece, no composite qubit may be involved in more than one composite gate. Since the smallest set of composite qubits has size $\min_j m_j$, this is how many composite gates we can fit into one piece. Thus, $\prod_{j=0}^hm_j/\min_j m_j$, as we wanted to show.
\hfill$\blacksquare$

Whether the number of pieces used by Theorem~\ref{thm3} is optimal is a question that is maybe best answered for specific codes. See for example the improvements over Theorem~\ref{thm3} that we achieved for the 7-qubit code in Section~\ref{resource_compare}. In addition, even ordering the $\text{C}^h\text{Z}$ gates differently within a piece changes the set of possible errors at an intermediate point. In general, any code will have unused syndromes at an intermediate error-correction (such as we noted for the 5-qubit code CZ), which may be exploitable through clever, code-specific circuit design.

It is interesting that, generically, the number of pieces required to implement a pieceably 1-fault-tolerant $\Gamma$-gate will not depend on the distance of the code (although the circuit size will). For instance, Theorem \ref{thm3} implies that, if we can use lowest weight normalizers to wire up the round-robin circuit (i.e. $w_j=d_j$ for all $j$), then implementing a $h+1$ qubit $\Gamma$-gate uses only $2^h$ pieces, independent of any $d_j$. Intuitively, this is because a large distance $d$ code, although requiring $d^{h+1}$ physical $\Gamma$-gates, also allows us to squeeze more gates into a single piece, while still guaranteeing correctable errors.

Finally, we note that the fact that each code $L_j$ encodes a single qubit is not necessary in the proofs of Theorems~\ref{thm1} and \ref{thm3}. The same arguments go through for doing $\Gamma$-gates between any $h+1$ encoded qubits, taking (your choice of) one encoded qubit from each $k\ge1$ code block. Generally, however, because we are not considering gates between encoded qubits in the same block anyway, we prefer to simplify to the $k=1$ case.

\subsection{Universality through pieceable means}
We now use the theorems from the previous section to construct universal sets of fault-tolerant logical gates on nondegenerate stabilizer codes. To do so, we have to prove facts about the constant stabilizers of stabilizer codes. The main technical result, proven in Appendix A, is the following lemma.
\begin{lem}\label{lem_ec}
Given a nondegenerate $\llbracket n,1,d\rrbracket$ stabilizer code with $d\ge3$ and stabilizer $S$, if $p\in\mathcal{N}(S)\setminus S$ has $|p|=d$, then $S_C(p)$ is error-correcting.
\end{lem}

It follows that, since all stabilizer codes have a normalizer with weight $d$, all nondegenerate stabilizer codes have a normalizer with an error-correcting constant stabilizer. Therefore, using Theorem~\ref{thm3}, we have that any nondegenerate stabilizer code with $d\ge3$ has pieceably fault-tolerant circuits for at least one of $\Gamma(ZZZ), \Gamma(XXX), \Gamma(YYY)$. With appropriate single-qubit gates, along with state preparation and measurement circuits we know exist for any stabilizer code, any one of these can form a universal set. In fact, for each of the three $\Gamma$-gates, adding a single qubit Clifford gate that does not commute with the gate will suffice for universality. For example, $\Gamma(ZZZ)$ plus $H$ is universal \cite{Shi2002,Aharonov2003}, as is $\Gamma(YYY)$ plus $SHS$. The octahedral gates of Eq.~\eqref{octahedral_gate} commute with none of these three $\Gamma$-gates, and so always complete a universal set.

The upshot of the brief argument of the previous paragraph are the following corollaries.
\begin{cor}\label{cor1}
Any nondegenerate stabilizer code with a complete fault-tolerant set of logical local Clifford gates has a universal set of fault-tolerant logical gates.
\end{cor}\noindent
\begin{cor}\label{cor2}
Any nondegenerate linear stabilizer code with distance $d\ge3$ has a fault-tolerant and universal set of logical gates.
\end{cor}\noindent
Obviously, the assumptions of Corollary~\ref{cor1} are a bit stronger than necessary given our argument above, but with this statement we have removed reference to constant stabilizers. Corollary~\ref{cor2} follows from the fact that $\text{GF}(4)$-linear codes (like the 5-qubit code) always have transversal octahedral gates, such as $K$.

We have a similar logical progression for CSS codes. In this case, from Appendix A we have the lemma
\begin{lem}\label{lem_css}
Any nondegenerate $\llbracket n,1,d\rrbracket$ CSS code with $d\ge3$ and stabilizer $S$ has two normalizers $p_1,p_2\in\mathcal{N}(S)\setminus S$ with $S_C(p_1)$ and $S_C(p_2)$ both error-correcting and $\{p_1,p_2\}=0$.
\end{lem}\noindent
The corresponding corollary is
\begin{cor}\label{css_cor1}
Any nondegenerate CSS code with a fault-tolerant logical local Clifford gate has a universal set of fault-tolerant logical gates.
\end{cor}\noindent
This corollary is true because by Lemma~\ref{lem_css} nondegenerate CSS codes have pieceably fault-tolerant circuits for at least two of $\Gamma(ZZZ), \Gamma(XXX), \Gamma(YYY)$. Every single-qubit Clifford gate fails to commute with at least two of these three $\Gamma$-gates, and as we noted above Corollary~\ref{cor1} noncommutation suffices for universality. The conclusion of Corollary~\ref{css_cor1} would similarly hold true for any code satisfying the conclusion of Lemma~\ref{lem_css}.

Finally, we have a code-switching corollary.
\begin{cor}\label{cor_switch}
For any two nondegenerate stabilizer codes, each of which either (a) is CSS with a logical local Clifford gate or (b) possesses a complete fault-tolerant set of logical local Cliffords, there exists a logical SWAP between them.
\end{cor}\noindent
This is simply a result of having CX from one code to the other, in both directions. It is perhaps more interesting to note the implication of a scheme for doing universal computation with any nondegenerate CSS code $L$ through code-switching, employing an ancilla codeblock. Using Lemma~\ref{lem_css}, we agree to encode $L$ such that $p_1$ is a logical $Z$ operator and $p_2$ is logical $X$. Then, code-switching is possible between $L$ and any code indicated in Corollary~\ref{cor_switch}, including codes like the $7$-qubit and $15$-qubit which together provide a universal set of transversal gates. The same argument also works for any code satisfying the conclusion of Lemma~\ref{lem_css}, and can also work to interact qubits encoded in the same block by switching some to ancillary blocks first. These possibilities demonstrate some of the ability of pieceable fault-tolerance once functional ancillas are allowed.

Degenerate codes are conspicuously absent from our results in this section. However, it should be noted that Theorem~\ref{thm3} does not require nondegeneracy, it only requires error-correcting constant stabilizers (condition 2). Nondegeneracy simply makes it easy to prove the existence of such a constant stabilizer (Lemma \ref{lem_ec}), but the converse is not true. In fact, some degenerate codes do indeed have error-correcting constant stabilizers, such as Shor's 9-qubit code \cite{Shor1995} for some logical $Z$ operators. Moreover, we can even show (Appendix C) it is possible for pieceable constructs to exist despite violating the second condition of Theorem \ref{thm3}. Finally, it is also worth noting that because we are dealing with 1-fault-tolerance, an essentially distance three concept, that really nondegenerate in the lemmas and corollaries of this section means: no weight two stabilizers. This provides a strengthening of Lemmas~\ref{lem_ec} and \ref{lem_css} (proofs follow from Lemma~\ref{lem1} in Appendix A) from requiring a nondegenerate code to requiring a code without weight two stabilizers, though, again, considering only 1-fault-tolerance, these are essentially the same idea regardless.

\section{Conclusion}
Although we have developed fault-tolerant universal sets of logical gates for many stabilizer codes, there is more design space to explore. For instance, it is natural to ask questions such as, what other logical gates can be implemented in a pieceably fault-tolerant manner, with or without functional ancillas, or, what are the fewest functional ancillas necessary to implement a given gate pieceably fault-tolerantly in a given code? Some interesting target gates for these questions might be the controlled-Hadamard and controlled-phase gates, which are powerful gates from a universality perspective as well \cite{Kitaev1997}, or important single-qubit gates such as $T$ \cite{Boykin2000}. Similarly, universal gates between logical qubits encoded in the same code block are also worth studying.

Another direction to explore is in finding more codes that support universal sets of pieceable fault-tolerant gates. All of our results, namely Corollaries~\ref{cor1}, \ref{cor2} and \ref{css_cor1}, provide sufficient conditions for pieceable universality, but not necessarily necessary ones. We already raised the question of degenerate codes, and if there is always some pieceable $\Gamma$-gates that can be constructed on them, like for Shor's 9-qubit code. Of course, once functional ancillas are allowed, magic states can also be used, so perhaps this question would make the most sense with an ancilla restriction. 

Moreover, we have restricted ourselves to concatenated codes, an assumption it would be interesting, and important, to lift. When we think about non-concatenated code families, such as surface codes, the poor scaling of the depth of our pieceable circuits with respect to code distance $d$ (and, correspondingly, with code size) coupled with the fact that noncontagious errors may build up over the circuit (as they are not always corrected at intermediate error-corrections) suggests no threshold would exist.

A related extension is to consider more robust pieceable fault-tolerance in higher distance codes. We have built logical circuits that can recover from \emph{single} faults, regardless of the distance $d$ of the code, and through concatenation we achieve tolerance to an arbitrary number of faults. Still, if the distance of the base code were $d=2t+1$, it would be nice to design logical circuits that could, even before any concatenation, recover from $t$ faults, or, in other words, are $t$-fault-tolerant. For instance, our circuits on distance $d$, nondegenerate CSS codes can be pieced so that they are $t$-fault-tolerant, since in that case, using CSS PARSEC, $t$ contagious errors are intermediately correctable even without measuring the nonconstant stabilizers. However, can PARSEC, and in particular the concept of a reliability guarantee for measuring nonconstant stabilizers, generalize to provide $t$-fault-tolerant pieceable gates on other codes?

Finally, there is the question of efficiency. A complete comparison of pieceable fault-tolerance against other known means of achieving logical universality on several codes with respect to fault-tolerance threshold, ancilla qubits required, and circuit depth is warranted. Also, we expect that along with rigorous threshold calculations will come an inevitable wave of micro-optimizations to pieceable fault-tolerance as we have laid it out here. For instance, changing from a CAT state syndrome measurement scheme for intermediate error-correction to something like a Steane \cite{Steane1997,Steane1998} or Knill \cite{Knill2005} scheme (except with the added difficulty of non-Pauli measurements) is likely to have threshold increasing effects. We also cannot ignore the possibility of macro-optimizations --- for instance, replacing the round-robin circuit with a simpler, but still pieceably fault-tolerant, design customized for a particular code.

The authors would like to thank Andrew Cross, Graeme Smith, and John Smolin for insightful discussions about this work. Special thanks go to Cody Jones for suggesting the circuit in Eq.~\eqref{ccz_combo_7b} and David Poulin for pointing out \cite{Knill1996}. We gratefully acknowledge funding from NSF RQCC Project No. 1111337 and ARO quantum algorithms program. T.J.Y. is thankful for a National Defense Science and Engineering Graduate (NDSEG) fellowship. R.T. acknowledges the support of the Takenaka scholarship and Frank fellowship.

\bibliography{references}

\appendix

\section{Classical codes within quantum ones}
There is a useful description of the constant stabilizer $S_C(p)$ (see Eq.~\eqref{const_stab}) that comes along with an (efficient) algorithm for finding it. Recall that the stabilizer $S$ of an $\llbracket n,k,d\rrbracket$ quantum code can be written as a bifurcated binary matrix, \cite{Gottesman1997, Aaronson2004}
\begin{equation}
S=\left[A|B\right]
\end{equation}
with $n-k$ rows and $2n$ columns. The $j^\text{th}$ row corresponds to stabilizer generator $\widetilde Z_j$. The $k^\text{th}$ Pauli of $\widetilde Z_j$ is $I,X,Y,Z$ if and only if $(A_{jk},B_{jk})=(0,0),(1,0),(1,1),(0,1)$. The signs of the stabilizer generators, either $\pm1$, can be stored separately, but are unnecessary for our purposes. The normalizer $l$ can likewise be written as $[a|b]$, for (row) vectors $a$ and $b$ of length $n$.

Now consider two transformations of this representation. First, we will apply local Clifford gates to put $p$ into $Z$-form, $[\vec0|b']$. This causes $S$ to change correspondingly as well. Then, restrict $S$ and $p$ to the ``window'' $W=\text{supp}(p)$ by removing all columns from their binary matrix representations that correspond to qubits outside $W$. Now $p_W=[\vec0|\vec1]$ for length $|p|$ vectors of 0s and 1s, and $S_W=[A'|B']$. Since we can swap qubits within $W$ (corresponding to switching columns of $S_W$) and multiply stabilizer generators (corresponding to adding rows of $S_W$ modulo two) without changing $p_W$, we can perform Gaussian elimination so that
\begin{equation}\label{S_W}
S_W=\left[\begin{array}{cc|cc}I_r&D&B_1&B_2\\0&0&B_3&B_4\end{array}\right].
\end{equation}
Letting $r$ be the rank of $A'$, the heights of the blocks are $r$ and $n-k-r$ and the widths are $r,|p|-r,r,|p|-r$. So $I_r$ is the $r\times r$ identity matrix. Note $r\le|p|-1$, because the first $r$ rows of $S_W$ represent stabilizers that must commute with $p$, and the fact that $I_r$ has only one 1 per row implies $D$ must have an odd number of 1s per row, and thus its width cannot be zero. 

We remark offhand that, for codes encoding one qubit, if $p$ has minimum weight within its coset (i.e. $|p|=\min\{|p'|:p'\in pS\}$), then a stronger fact is true, namely $r=|p|-1$. This is so because all errors restricted to the support of $p$, consisting of only Pauli $Z$s and $I$s, and with weight $<|p|$ are detectable. Exactly half of these detectable errors must also be correctable, because, for any such error $E$, only $E$ and $pE$ will have the same syndrome. Thus,
\begin{equation}
\frac12\left(\binom{|p|}{1}+\binom{|p|}{2}+\dots+\binom{|p|}{|p|-1}\right)=2^{|p|-1}-1
\end{equation}
such errors are correctable. Only the first $r$ stabilizers in $S_W$ have the possibility to trigger on these errors. Thus, to have enough unique syndromes to correct them all, we must have $r\ge |p|-1$.

Continuing our discussion of the constant stabilizer $S_C(p)$, we now notice that $S_C(p)$ is generated by the stabilizers corresponding to the last $n-k-r$ rows of $S_W$. We observe that those same rows $H=\left[B_3B_4\right]$ represent the parity check matrix of a classical code. We call this the classical code $C(p)$ induced by $p$ on quantum code $L$. Not all rows of $H$ are necessarily linearly independent, but the essential parity checks can be recovered by removing linearly dependent rows. It is also worth mentioning that the first $r$ rows of $S_W$ correspond to a minimal generating set for the nonconstant stabilizers.

The error-correcting properties of $C(p)$ are directly related to the error-correcting properties of $S_C(p)$. Indeed, $S_C(p)$ is error-detecting if and only if $C(p)$ has classical code distance $d_{C(p)}\ge2$. Likewise, $S_C(p)$ is error-correcting if and only if $C(p)$ has classical code distance $d_C\ge3$. The classical errors detected or corrected, respectively, are exactly the contagious errors from $\mathcal{E}_C(p)$, Eq.~\eqref{cont_errs}.

We now show the following.
\begin{lem}\label{lem1}
Let $L$ be an $\llbracket n,1,d\rrbracket$ stabilizer code with stabilizer $S$ and $p\in\mathcal{N}(S)\setminus S$. 
\begin{enumerate}[(1)]
\item If $d\ge2$ and $|p|<2d-1$, then $C(p)$ exists and has distance $d_{C(p)}\ge2$.
\item If $p$ is minimum weight within its coset, there are no nonconstant stabilizers of weight two, and $C(p)$ exists with $d_{C(p)}\ge2$, then $d_{C(p)}\ge3$.
\item If $d\ge3$, $|p|=d$, and $L$ is nondegenerate, then $d_{C(p)}\ge3$.
\item If $d_{C(p)}\ge 3$, then there are no nonconstant stabilizers of weight two.
\item If $L$ is nondegenerate CSS and $l$ consists only of Pauli $Z$s and $I$s or only of $X$s and $I$s, then $d_{C(p)}\ge d$.
\end{enumerate}
\end{lem}\noindent
Given the correspondence between the error-correcting properties of $S_C(p)$ and $C(p)$ discussed above, it should be clear that part (3) of Lemma~\ref{lem1} results directly in the proof of Lemma~\ref{lem_ec}. Part (5) of Lemma~\ref{lem1} implies Lemma~\ref{lem_css} after one notices that, for any CSS code, there is always a normalizer in $Z$-form (only $Z$s and $I$s) and another in $X$-form (only $X$s and $I$s), and these belong to different cosets of the stabilizer \cite{Wilde2009}. Part (4) of Lemma~\ref{lem1} is used in the proof of Theorem~\ref{thm3}.

\emph{Proof of Lemma~\ref{lem1}}. Note that (3) is merely a consequence of (1) and (2). To prove the first two, we actually show something stronger: if $d\ge2$ and for all $l\in\mathcal{N}(S)\setminus S$ we have $d_{C(l)}\le2$, then there is a weight two stabilizer. The argument here proceeds recursively --- we either find a weight two stabilizer or we find a normalizer with smaller weight than we began with. At the bottom of the recursion (i.e. $|p|<2d-1$ or $p$ minimum weight within its coset) we will get (1) and (2).

We work with the windowed stabilizer, Eq.~\eqref{S_W}, windowed on $l\in\mathcal{N}(S)\setminus S$. If $H=[B_3B_4]$ does not exist (i.e. $r=n-k$), then we choose a column from $B=[B_1B_2]$, say column $k$, and define $\mathcal{B}=\{j:B_{jk}=1\}$. Then $e=\prod_{j\in\mathcal{B}}Z_jX_k$, whose weight notice is $|e|\le|l|$, commutes with all stabilizers but anticommutes with $l$. Therefore, $e\in\mathcal{N}(S)\setminus S$ and $el\in\mathcal{N}(S)\setminus S$. So, $|e|\ge d$ and $|l|-|e|+1\ge d$ together imply $d\le|e|\le|l|-d+1$. Thus, because $d\ge2$, $|e|\le|l|-1$, and we found a lower weight normalizer.

If $H=[B_3B_4]$ does exist, then it either has a column of 0s (distance $d_{C(l)}=1$) or has two identical columns (distance $d_{C(l)}=2$). In the former case, let the column of 0s be $k$ and define $\mathcal{B}'=\{j:B_{jk}=1\}$ and $e=\prod_{j\in\mathcal{B}'}Z_jX_k$. The argument proceeds as before, so we omit it. But in the latter case, say the identical columns are $k_1$ and $k_2$. Define $\mathcal{B}_{12}=\{j:B_{jk_1}\neq B_{jk_2}\}$ and $e=\prod_{j\in\mathcal{B}_{12}}Z_jX_{k_1}X_{k_2}$. Note $|e|\le|l|$ and additionally $e$ commutes with all stabilizers and $l$, so $e\in lS$ or $e\in S$. In the former case, if $|e|=|l|$, then $el\in S$ has weight two. In the latter case, $el\in lS$ and $|el|\le|l|$. If $|el|=|l|$, then $\mathcal{B}_{12}\subseteq\{k_1,k_2\}$, implying $|e|=2$.

The assumptions of (1) mean that if we started this recursive process with $l=p$, we would only be able to conclude that $H$ has no column of 0s and thus $d_{C(p)}\ge2$. Likewise, if we started with the assumptions of (2) instead, the conclusion would be $d_{C(p)}\ge3$.

Part (4) is simple to prove in contrapositive. If there is a nonconstant stabilizer of weight two, it must have the form $X_{k_1}X_{k_2}$, $X_{k_1}Y_{k_2}$, $Y_{k_1}X_{k_2}$, or $Y_{k_1}Y_{k_2}$ on two qubits $k_1,k_2\in\text{supp}(p)$. Since $X$ and $Y$ are the same from the perspective of commutation with constant stabilizers, we can consider just the first case. In that case, the constant stabilizers cannot distinguish between $X_{k_1}$ and $X_{k_2}$, which means $C(p)$ is not error-correcting.

Finally, we finish with the last part (5). This one is comparatively simple. Recall the stabilizer of a CSS code can always be written like \cite{Calderbank1996,Steane1996b}
\begin{equation}
S=\left[\begin{array}{c|c}H_1&0\\0&H_2\end{array}\right].
\end{equation}
Thus, if $p$ is in $Z$-form (consists only of $Z$s and $I$s), then $H_2$ is the check matrix of the classical code $C(p)$. But since the quantum code has distance $d$ and is nondegenerate, this classical code must have distance at least $d$ as well. Likewise, if $p$ is in $X$-form, then $H_1$ is the check matrix of $C(p)$, which must also have distance at least $d$.
\hfill$\blacksquare$

\section{Nonstabilizer intermediate codes}
Here we show that the codes we encounter intermediate in the round-robin constructions of Theorem~\ref{thm1} are in fact nonstabilizer codes. This implies that doing complete error-correction, rather than just correcting contagious errors as we do with PARSEC, will likely require the discovery of new tools. Only a few examples are known of doing error-correction on non-stabilizer codes, such as CWS codes \cite{Li2010}.

In particular we prove the following:
\begin{lem}\label{lem_app}
Assume that we use normalizers with weight minimal within their coset to construct logical $\Gamma$-gates with $h\ge2$ in Theorem~\ref{thm1}. At any point in the round-robin circuit, if the codespace is not equal to the codespace at the beginning of the round-robin circuit, the codespace is non-stabilizer.
\end{lem}\noindent
\emph{Proof.} Assume we have converted each of the $h+1$ code blocks into $Z$-form. We take $h\ge2$ because our conclusion is obviously false for round-robin CZ constructions. Then the stabilizer of any one code block, before any gates of the round-robin are applied takes the form of $S_j=\langle t_{j1},\dots,t_{jr},s_{j1},\dots,s_{jm}\rangle$ where $m=n-1$. We have written nonconstant stabilizer generators as $t$ and constant stabilizer generators as $s$. We assume that we have found a maximal set of constant stabilizers (e.g.~using the methods of Appendix A). Also, let $\overline{p}_j$ denote the $Z$-form normalizer of code $j$, the one with minimal weight.

Let $S$ denote the complete stabilizer of all $h+1$ codes before any round-robin gates have been applied. The transformation of $S$ after some number of gates, a subcircuit $C$, will be denoted $S'$, and the transformation of $S_j$ as $S_j'$. Notice that $t_{jk}'=t_{jk}G_{jk}$, where $G_{jk}$ is a product of $\text{C}^{h-1}\text{Z}$ gates on the qubits in code blocks other than block $j$. However, this is the extent of the transformation of the stabilizer --- the constant stabilizers are preserved, $s_{jk}'=s_{jk}$. 

There are three properties of $G_{jk}$ to note: (1) $G_{jk}^2=I$ (2) $G_{jk}=\frac{1}{M_{jk}}\sum_{m}g^m_{jk}$, where $g^m_{jk}$ are Pauli operators (with $\pm1$ signs) consisting of only $Z$s and $I$s supported on $\bigcup_j \text{supp}(\overline{p}_j)$ and $M_{jk}>0$ is normalization (3) $[G_{jk},s_{uv}]=0$ for all $j,k,u,v$.

Let $P'$ denote the projector onto the codespace after $C$. Explicitly,
\begin{equation}
P'=\frac{1}{N_S}\sum_{g\in S'}g=\frac{1}{N_S}\sum_{g\in S}CgC^\dag,
\end{equation}
where $N_S$ is just normalization so that $P'^2=P'$. We want to show that there is no subgroup $T$ of the Pauli group such that
\begin{equation}
P'\propto\sum_{g\in T}g.
\end{equation}
To prove $P'$ is not of this stabilizer form, it is sufficient to show that two terms in the Pauli decomposition of $P'$ have coefficients that differ in magnitude. Since the constant stabilizers are part of the Pauli decomposition of $P'$ and their coefficients have unit magnitude, we only need to find a Pauli in the decomposition with a coefficient of magnitude less than one.

If all $G_{jk}$ are identity we are in the original codespace. Therefore, one of $G_{jk}$ is not identity. Now, $G_{jk}$ cannot be a single Pauli term $G_{jk}=g^0_{jk}$. This is because $g^0_{jk}$, as a non-identity Pauli consisting of $Z$s and $I$s, has some $-1$ eigenstate $|b\rangle$ with $b$ a weight one bit string. However, for $G_{jk}$, as a product of $\text{C}^{h-1}\text{Z}$ gates with $h\ge2$, all eigenstates of this form have eigenvalue $+1$. Since $G_{jk}$ consists of at least two different terms then $1/M_{jk}$, the normalization of $G_{jk}$, must be at most $1/\sqrt{2}$.

We now argue that, for all $m$, $t_{jk}g^m_{jk}$ appears only once in the Pauli decomposition of $P'$, namely as part of the decomposition of $t_{jk}G_{jk}$. If no other $g^l_{jk}\neq g^m_{jk}$ exists such that $g^l_{jk}g^m_{jk}$ is a stabilizer, this is certainly true. After all, $t_{jk}$ as a nonconstant stabilizer can only appear on block $j$ from one combination of stabilizer generators, namely from $t_{jk}G_{jk}$ itself. Indeed, it cannot be the case that $g^l_{jk}g^m_{jk}$ is a stabilizer, as this would imply (since both $g^l_{jk}$ and $g^m_{jk}$ are supported only in $\bigcup_j \text{supp}(\overline{p}_j)$ and only by $Z$s) a constant stabilizer exists in at least one block $i\neq j$ that could be multiplied by the normalizer $\overline{p}_i$ to get a lower weight normalizer.
\hfill$\blacksquare$

This lemma can be used to verify that an intermediate codespace is non-stabilizer. In particular, it says, in the case of using lowest weight normalizers for the construction, that if any nonconstant stabilizer $t_{jk}$ takes the intermediate form $t_{jk}G_{jk}$, where $G_{jk}$ is a product of $\text{C}^{h-1}\text{Z}$ gates not equal to identity, then the codespace is non-stabilizer.

\section{Demonstrating pieceability beyond Theorem \ref{thm3}}
In this appendix, we address the question of whether condition (2) of Theorem \ref{thm3} is necessary for a fault-tolerant pieceable circuit implementing logical $\Gamma(p)$ to exist. The answer is ``no". To show this, we provide a simple example implementing CZ on Shor's 9-qubit code. Thus, codes and gates $\Gamma(p)$ satisfying the conditions of Theorem \ref{thm3}, such as the 5-qubit and 7-qubit codes in the main text, are put in context as simply convenient examples of pieceable fault-tolerance, rather than the extent of its applicability.

Recall that Shor's 9-qubit code \cite{Shor1995} is the concatenation of two redundancy codes, one $L_Z$ correcting phase flip errors and one $L_X$ correcting bit flips. We will take $L_Z$ as the outer code, thus forming the stabilizer as
\begin{equation}
S_9=\bigg\langle\begin{array}{ccc}X_1X_2X_3X_4X_5X_6&Z_1Z_2&Z_2Z_3\\X_4X_5X_6X_7X_8X_9&Z_4Z_5&Z_5Z_6\\&Z_7Z_8&Z_8Z_9\end{array}\bigg\rangle.
\end{equation}
Logical operators are taken as $\overline{Z}_9=X^{\otimes9}$ and $\overline{X}_9=Z^{\otimes9}$, defined this way according to convention \cite{Nielsen2010}. Shor's code is CSS so it possesses a transversal $\Gamma(ZX)=\text{CX}$ gate.

\begin{figure}[t]
\includegraphics[width=\columnwidth]{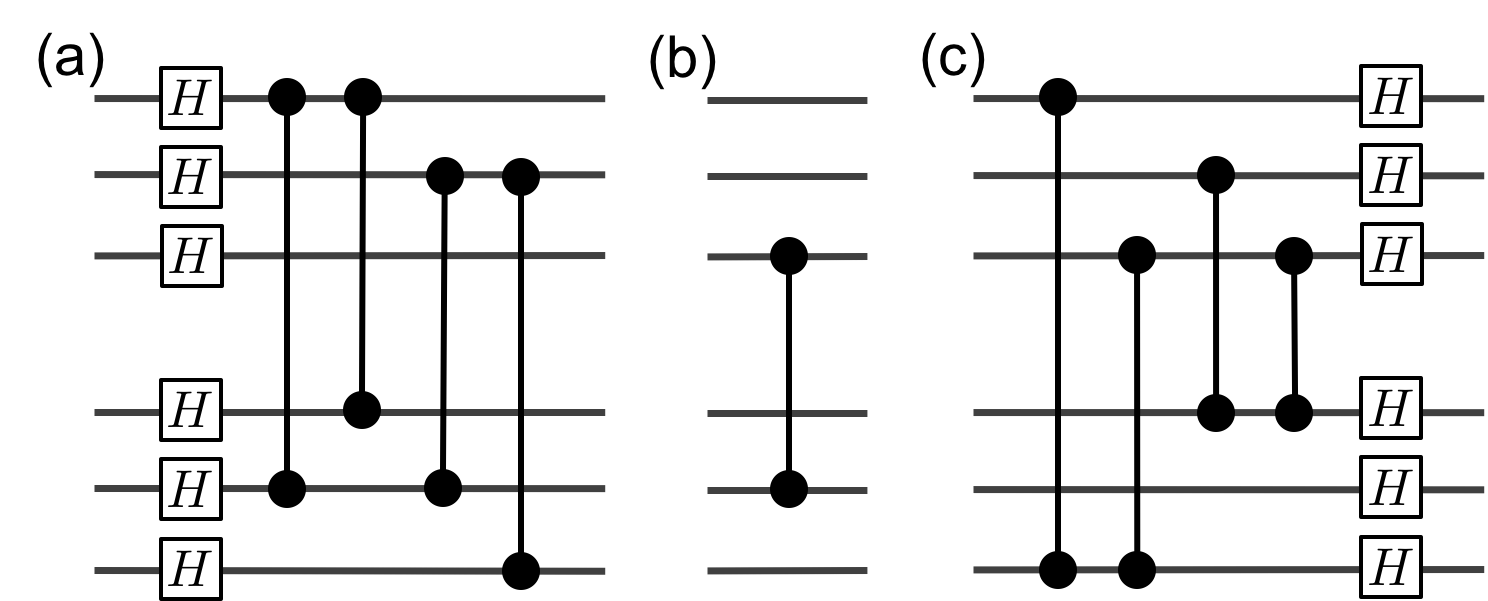}
\caption{\label{fig5} A pieceable implementation of $\text{CZ}$ on Shor's 9-qubit code using three pieces. Only active qubits, namely $1,2,3$ from each code block, are shown. Since the circuit is Clifford, intermediate error-correction consists of standard Pauli measurements.}
\end{figure}

Using a computer and the methods of Appendix A, it is easy to check that, while Shor's code does possess logical $X$ operators with error-correcting constant stabilizers (such as $Z_1Z_4Z_7$), no logical $Z$ operator has an error-correcting constant stabilizer. This makes a pieceable $\Gamma(XX)$ gate (or $\Gamma(XXX)$ for that matter) implementable by Theorem \ref{thm3}, but not a $\Gamma(ZZ)=\text{CZ}$ gate. Gauge-fixing can implement a logical Hadamard on Shor's code \cite{Aliferis2007}, and thus implement CZ from CX (not to mention completing a universal gate set when placed alongside pieceable $\Gamma(XXX)$). But the question remains whether gauge-fixing is necessary for a logical CZ, or if pieceable alone is enough.

It turns out CZ is implementable in pieceable fashion, as shown in Fig.~\ref{fig5}. Only the active qubits are shown, which can be taken as qubits $1,2,3$ from each block. This pieceable circuit is clearly the round-robin circuit that implements logical $\text{CZ}$ according to Theorem \ref{thm1}. However, it is not and cannot be pieced as specified by Theorem \ref{thm3} while remaining fault-tolerant. In fact, even the order of CZ gates within a piece is important to the success of the intermediate error-corrections. By exhaustive computer search over permutations of the CZ gates and placement of the intermediate error-corrections, three is the smallest number of pieces making this round-robin circuit on Shor's code fault-tolerant.

\section{Circuits for resource comparison}
In Section~\ref{resource_compare}, we counted ancilla qubits, mult-qubit gates, and circuit volume for two implementations of logical CCZ on the 7-qubit code -- magic state injection via postselection and pieceable fault-tolerance. Here we draw the circuits we used in both cases. The captions explain their circuit volumes. The number of ancilla qubits and number of CX and CCZ gates involved are also easily countable from the diagrams.

Of course, there are potential improvements and tradeoffs in both circuit designs. For instance, reducing the size of the CAT states in the magic state preparation from 4 qubits to 2 is possible but will lead to increased circuit volume, as coupling to the $\ket{\overline{\text{CCZ}}}$ register will take more additional volume than is removed by preparing a smaller CAT.
However, such optimizations are unlikely to erase the nearly 50\% improvement (see Table~\ref{tab1}) that a pieceable CCZ offers.

\begin{figure*}[b]
\includegraphics[width=1.8\columnwidth]{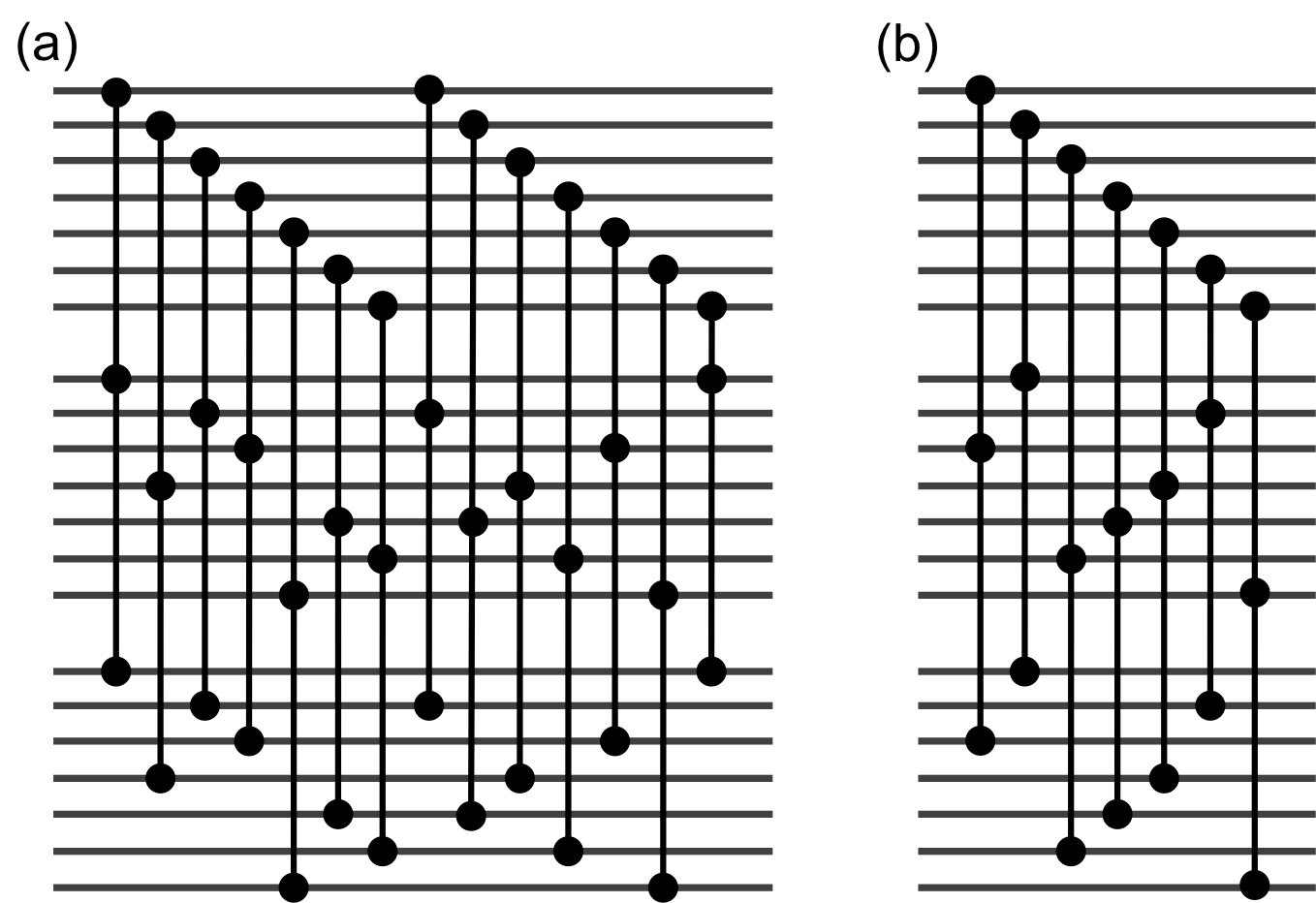}
\caption{\label{shortest_ccz} A pieceable circuit for logical CCZ on the 7-qubit code that uses just 21 physical CCZ gates and two pieces. Intermediate error-correction consists of Steane syndrome measurement of the $Z$-stabilizers on each block. The final error-correction is full Steane syndrome measurement on all blocks. We assume that recovery need not be applied directly. Instead, $X$-errors detected intermediately can be dealt with by adaptively changing the following CCZ gates to be $|0\rangle$-controlled on the affected qubits. The first piece has volume $2\times21=42$ and the second just $21$. Using the circuit volumes of the measurement circuits in Fig.~\ref{steane_measure}, we find the total volume of the pieceable CCZ to be $3\times21+3\times81+3\times155=771$.}
\end{figure*}

\begin{figure}
\includegraphics[width=\columnwidth]{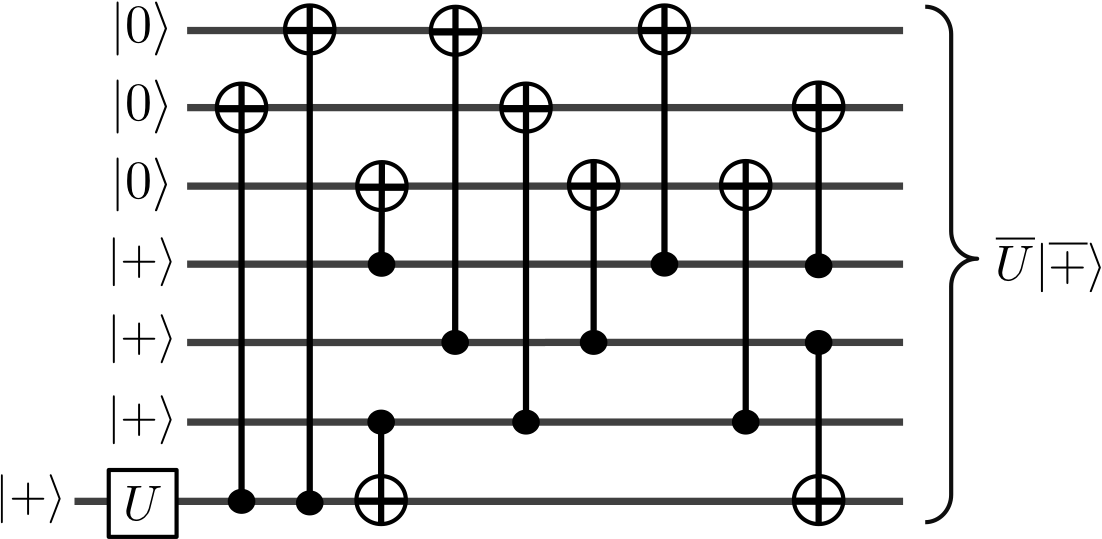}
\caption{\label{encode_7qubit} The non-fault-tolerant circuit, based off the design in \cite{Goto2016}, for encoding a single-qubit state into the 7-qubit code. Pushing state preparations as far right as possible, we find that this construction has circuit volume 34. To make $|\overline{\text{CCZ}}\rangle$ (assuming no faults) we can use three such circuits on three physical qubits in the state $\ket{\text{CCZ}}=\text{CCZ}|+\rangle^{\otimes3}$.}
\end{figure}

\begin{figure}
\includegraphics[width=\columnwidth]{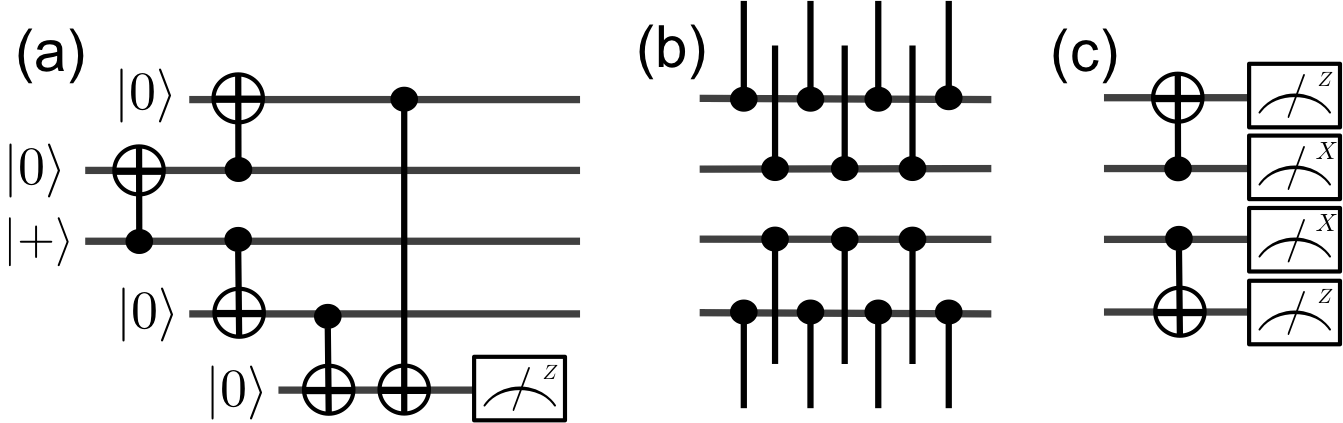}
\caption{\label{4CAT_plus_decode} (a) Preparing and verifying $\text{4-CAT}=(|0000\rangle+|1111\rangle)/\sqrt{2}$ (the measurement is postselected on finding $|0\rangle$). The circuit volume is 26 and we must wait for the postselection to succeed before proceeding. (b) The circuit coupling the 4-CAT to the $|\overline{\text{CCZ}}\rangle$ blocks in Fig.~\ref{ccz_state_postselect}. All outgoing lines are to unique qubits, so this takes 4 timesteps for a total circuit volume of 16. (c) The CAT decoding \cite{Divincenzo2007} and measurement step. Successful postselection occurs upon seeing $\ket{0}$ from both $Z$-basis measurements. The xor of the two $X$-basis measurements gives the eigenvalue of the measured operator. The volume of this step is 8. Total volume of (a),(b),(c) is 50.}
\end{figure}

\begin{figure}
\includegraphics[width=\columnwidth]{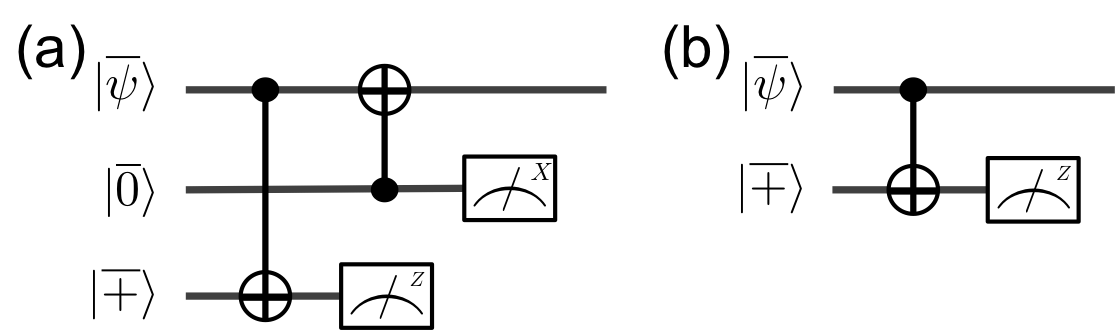}
\caption{\label{steane_measure} (a) Measurement of the stabilizers of the 7-qubit code block $\ket{\overline\psi}$ as put forward by Steane \cite{Steane1997}. Gates and measurements are drawn to represent transversal versions. Both $\ket{\overline0}$ and $\ket{\overline{+}}$ are verified logical states that we assume are created by Goto's postselection \cite{Goto2016} circuit with just 1-qubit for verification (so 8 qubits total), which has circuit volume 53. The total circuit volume of the complete Steane stabilizer measurement is $53\times2+49=155$. (b) The measurement of just the $Z$-stabilizers on the 7-qubit code using a verified $\ket{\overline{+}}$ state. Total circuit volume is $81$.}
\end{figure}

\begin{figure}
\includegraphics[width=\columnwidth]{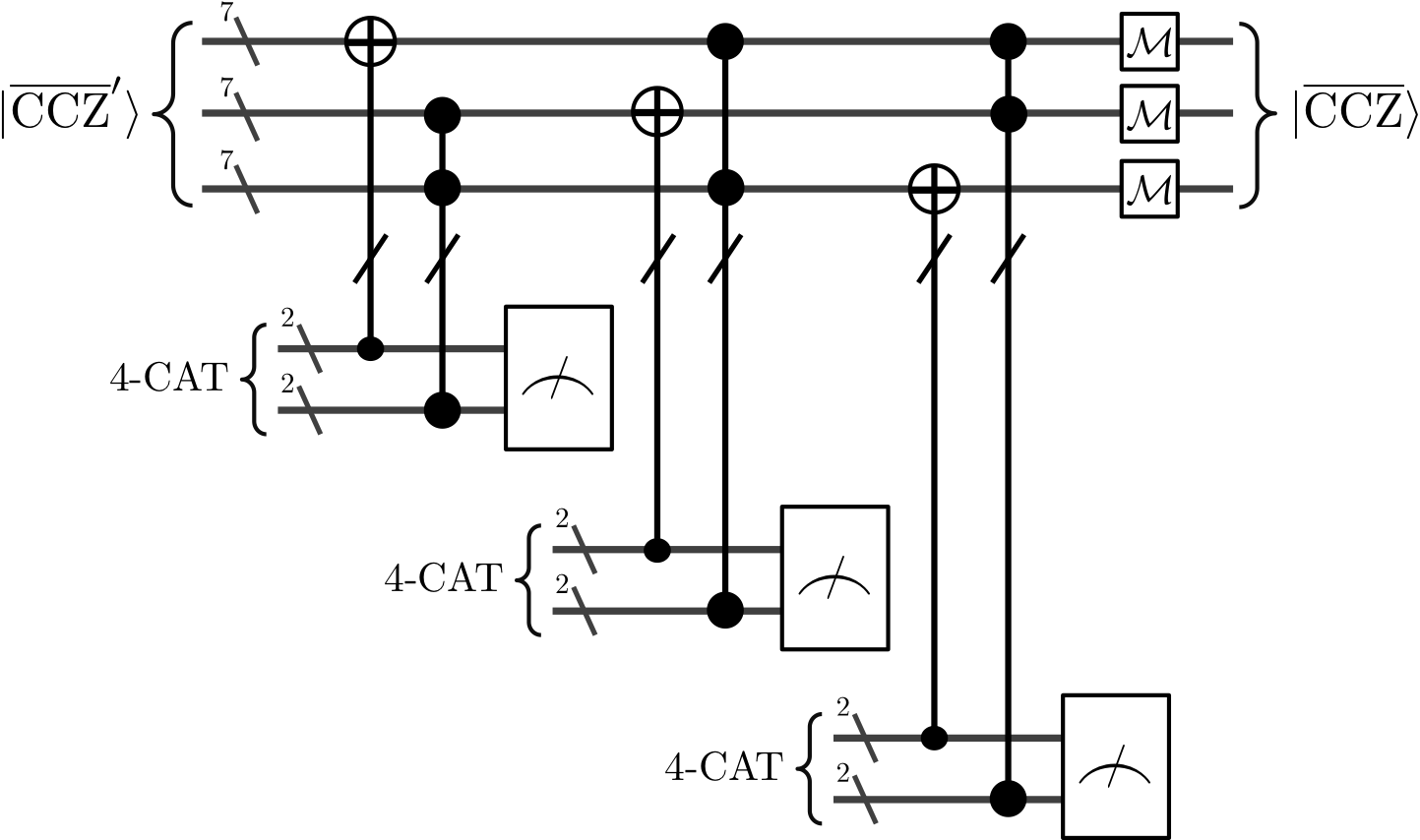}
\caption{\label{ccz_state_postselect} The postselected verification of the potentially faulty $|\overline{\text{CCZ}}'\rangle$ state that is output from the non-fault-tolerant preparation in Fig.~\ref{encode_7qubit}. The slashed CX and CCZ gates are controlled on the 4-CATs as shown in Fig.~\ref{4CAT_plus_decode}(b) and targeted as transversal $X$ and transversal CZ on the encoded blocks. The 4-CAT inputs are prepared as in Fig.~\ref{4CAT_plus_decode}(a) and the measurement boxes on the 4-CATs represent Fig.~\ref{4CAT_plus_decode}(c). We use $\mathcal{M}$ to represent measurement of all 7-qubit code stabilizers using Steane states, Fig.~\ref{steane_measure}(a). In this case, a successful postselection occurs upon a trivial syndrome from all $\mathcal{M}$ and from the CAT measurements. The total circuit volume is $34\times3+50\times3+4\times21\times3+155\times3=969$.}
\end{figure}

\begin{figure}
\includegraphics[width=\columnwidth]{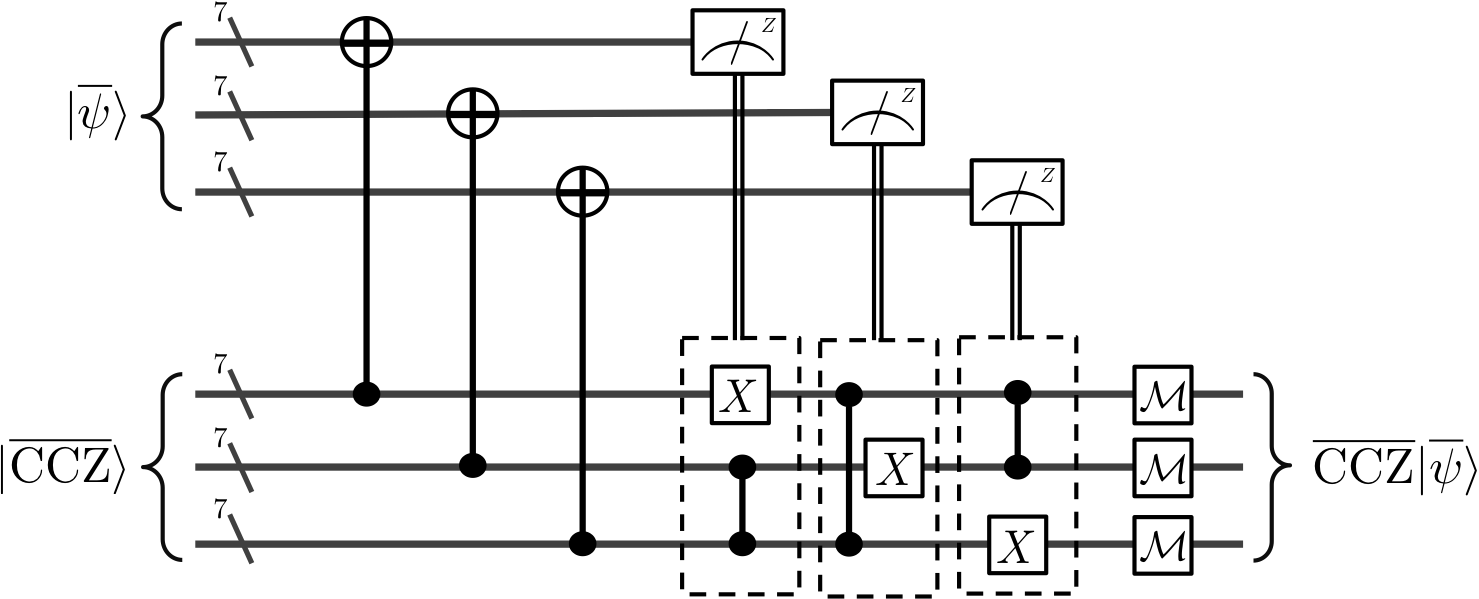}
\caption{\label{ccz_teleported} Teleporting a CCZ gate into code blocks using a $|\overline{\text{CCZ}}\rangle$ state verified using Fig.~\ref{ccz_state_postselect}. All gates and measurements are transversal. Again, $\mathcal{M}$ is the measurement of all 7-qubit stabilizers using Steane states, Fig.~\ref{steane_measure}(a). Assuming the classically controlled operations are never done, the total circuit volume, the complete volume of magic state injection with postselected preparation, is $969+42\times2+155\times3=1518$.}
\end{figure}

\end{document}